\documentclass[twocolumn]{aastex631}

\usepackage{booktabs}
\usepackage{float}
\usepackage{graphicx}

\begin{document}

\title{A Framework for Obtaining Accurate Posteriors of Strong Gravitational Lensing Parameters with Flexible Priors and Implicit Likelihoods using Density Estimation}

\author{ \href{https://orcid.org/
0000-0001-9459-6316}{Ronan Legin}}
\affiliation{Department of Physics, Universit\'{e} de Montr\'{e}al, Montr\'{e}al, Canada}

\author{\href{https://orcid.org/0000-0002-8669-5733}{Yashar Hezaveh}}
\affiliation{Department of Physics, Universit\'{e} de Montr\'{e}al, Montr\'{e}al, Canada}
\affiliation{Center for Computational Astrophysics, Flatiron Institute, 162 5th Avenue, 10010, New York, NY, USA}

\author{ \href{https://orcid.org/0000-0003-3544-3939}{Laurence Perreault-Levasseur}}
\affiliation{Department of Physics, Universit\'{e} de Montr\'{e}al, Montr\'{e}al, Canada}
\affiliation{Center for Computational Astrophysics, Flatiron Institute, 162 5th Avenue, 10010, New York, NY, USA}
\affiliation{Mila - Quebec Artificial Intelligence Institute, Montr\'{e}al, Canada}

\author{\href{https://orcid.org/
0000-0002-5854-8269}{Benjamin Wandelt}}
\affiliation{Center for Computational Astrophysics, Flatiron Institute, 162 5th Avenue, 10010, New York, NY, USA}
\affiliation{Institut Lagrange de Paris, Sorbonne Universit\'{e}, Paris, France}



\begin{abstract}

We report the application of implicit likelihood inference to the prediction of the macro-parameters of strong lensing systems with neural networks. This allows us to perform deep learning analysis of lensing systems within a well-defined Bayesian statistical framework to explicitly impose desired priors on lensing variables, to obtain accurate posteriors, and to guarantee convergence to the optimal posterior in the limit of perfect performance. We train neural networks to perform a regression task to produce point estimates of lensing parameters. We then interpret these estimates as compressed statistics in our inference setup and model their likelihood function using mixture density networks. We compare our results with those of approximate Bayesian neural networks, discuss their significance, and point to future directions. Based on a test set of 100,000 strong lensing simulations, our amortized model produces accurate posteriors for any arbitrary confidence interval, with a maximum percentage deviation of 1.4\% at 21.8\% confidence level, without the need for any added calibration procedure. In total, inferring 100,000 different posteriors takes a day on a single GPU, showing that the method scales well to the thousands of lenses expected to be discovered by upcoming sky surveys.

\end{abstract}



\section{Introduction} \label{intro}

Strong gravitational lensing is a natural phenomenon in which the light rays of distant galaxies are deflected by the gravity of foreground matter, resulting in the production of multiple images. It is a powerful probe that can map the inner distribution of matter in individual lens galaxies to reveal invaluable information about the physics of dark matter  \citep[e.g., ][]{2002ApJ...572...25D, 2012Natur.481..341V, 2016ApJ...823...37H}. It can also provide precise estimates of the expansion rate of the Universe, the Hubble constant \cite[e.g.,][]{1997ApJ...487...42K, 2020MNRAS.494.6072S,  2020MNRAS.498.1420W}. Strong lensing can also be used to study the magnified images of background galaxies, which are typically some of the most distant galaxies of the Universe, effectively working as a natural telescope \cite[e.g.,][]{2012ApJ...747L...9Z, 2013ApJ...762...32C, 2019ApJ...870L..11F}. 

Achieving all these science goals with strong gravitational lensing requires knowledge of lensing distortions, a process commonly referred to as lens modeling, in which the true, unlensed image of background sources and the physical distribution of matter in lensing structures are reconstructed. The parameters defining these physical components (e.g., the ellipticity of lensing galaxies) are generally referred to as ``lens parameters'', and they represent latent variables for which an estimate is desired. Traditionally, this inference has been done using density samplers (typically a Markov Chain Monte Carlo, MCMC) to map an explicit posterior over the lens parameters. In this context, the models are simulations of strong lenses with parameters proposed by MCMCs. This is, however, a time and resource-consuming process and is difficult to automate. In addition, it requires explicit closed-form posteriors, which, for real data, is often only possible as an approximation to the true posterior (e.g., when residuals from subtracting the main galaxy deflector light are present over the strong lensing arcs), potentially resulting in biased estimates.

In the coming years, a new generation of sky surveys, in particular, the Legacy Survey of Space and Time at Vera Rubin Observatory (LSST, 2023) and the Euclid Space Telescope (2022), are poised to discover more than 200,000 new strong gravitational lenses \citep{2015ApJ...811...20C}. Estimating the lens parameters with these traditional modeling approaches is intractable for the monumental volumes of data from these upcoming surveys.

In recent years, machine learning has provided a different alternative. By training convolutional neural networks to perform a regression task, it was shown that it is possible to obtain point estimates of lens parameters extremely accurately more than 10 million times faster than traditional methods \citep{2017Natur.548..555H}. Subsequent works have expanded this result to also obtain uncertainty estimates for the predictions made by neural networks. This has been done using approximate Bayesian neural networks trained with variational inference \citep{2017ApJ...850L...7P, 2019ApJ...883...14M, 2022arXiv220300690W, 2021ApJ...909..187W, 2021ApJ...910...39P}. These procedures can provide interval estimates that need to be calibrated in a frequentist manner (e.g., with coverage probabilities). Despite their success at producing accurate measurements in controlled experiments, these procedures involve many levels of approximations (e.g., the choice of the variational distributions for the outputs and the network weights), whose effect cannot be easily quantified or controlled, possibly resulting in biased estimates (e.g., when modeling multi-modal distributions, \citealp{2020arXiv200601490C}). Additionally, despite their name, they do not offer a truly Bayesian inference framework for the lens parameters: there is no clear way to impose explicit priors on them. In theory, the distribution of parameters in the training sets could be interpreted as the prior. However, given the limited expressiveness of realistic models, this also interferes with the learning and the performance of the networks over improbable regions of the prior. The fact that rigorous uncertainty quantification is crucial for the science goals enabled by strong lensing suggests that more attention should be given to the exact way in which this inference problem is formulated. 

This is fortunately possible, thanks to various implicit likelihood inference methods, also commonly referred to as likelihood-free inference or simulation-based inference. In this framework, instead of explicitly writing a noise model and sampling from it, repeated simulations of the physical model, including stochastic processes and noise, are used to obtain a Monte Carlo approximation of the posterior. In the limit of accurate simulations of the full data set, this is guaranteed to converge to the true, optimal posterior. For high-dimensional data sets, however, this can be computationally intractable. Instead, many methods resort to using compressed statistics of the data to facilitate computations. If the compressed representations are sufficient statistics for the parameters of interest, the obtained posteriors remain unchanged. If these compressions have biases and/or lose pertinent information, the obtained posteriors will be less constraining, however, they will still produce an accurate estimation of the posterior given the retained information. With the advent of machine learning, numerous methods have used these techniques to accelerate and improve these likelihood-free inference methods \citep{2019arXiv191101429C}. Recent works have demonstrated the application of these models to the estimation of low-dimensional (e.g., 1-2) parameters in lensing problems \citep{2019ApJ...886...49B, 2020arXiv201007032C}. 

In this work, we show that by using implicit likelihood inference methods, it is possible to perform accurate inference of strong gravitational lensing posterior distributions for thousands of systems using neural networks. Similar to \cite{2017Natur.548..555H}, we first train a convolutional neural network with a mean squared error loss to produce point estimates of the lensing parameters. These parameters describe both the background source (S\'{e}rsic) and the foreground lens (Singular Isothermal Ellipsoid, SIE, with external shear). Once the networks are trained, we generate 100 million simulations from the prior distribution, pass them to this network, and obtain their predictions. We then model the distribution of true vs. predicted parameters by training a density estimation model (a mixture density network) to learn the likelihood function of the learned parameters (compressed statistics). At inference time, the posterior distributions over lensing parameters conditioned on the compressed data are sampled using an MCMC from the density network and using arbitrary priors. We demonstrate the accuracy of these posteriors and discuss them in detail. 

In section \ref{methods}, we describe our lensing simulations and explain the motivation for using density estimation methods. In section \ref{results_tests}, we detail the tests used to evaluate the accuracy of the predicted posteriors along with their results. In section \ref{discussion}, we discuss the performance of this method in inferring accurate lensing posterior distributions and point to future applications.

\section{Methods} \label{methods}

\subsection{Implicit Likelihood Inference}

Traditionally, the analysis of astrophysical data is performed using statistical models, commonly within a Bayesian framework. In this context, simulated data sets, generally called “models”, are produced and compared to the data in order to measure latent variables, generally called “parameters” in astrophysics. The production of simulated data (e.g., pixel values for images) is controlled by model parameters (e.g., the mass of a lensing galaxy in the simulation), and the comparison to the data is made via a likelihood function. Using Bayes' theorem, one can then obtain a probability distribution (the posterior) over all possible parameter values.
\begin{equation}
    p(\theta | x) = \frac{p(x | \theta) p(\theta)}{p(x)},
\end{equation}
where $p(\theta | x)$ is the posterior distribution of model parameters $\theta$ given observed data $x$, and $p(x | \theta)$, $p(\theta)$, and $p(x)$, are the likelihood, prior, and evidence distributions, respectively.

Bayes' theorem states that the probability of the model (and consequently its parameters) given the current data is proportional to the probability of producing the data given the current model. In other words, one has to be able to calculate how likely it is to observe the current data if the model under consideration is indeed the underlying truth. This is given by the likelihood function, $p(x | \theta)$. After multiplying this by the prior, the posterior is then mapped using samplers, for example, with a Markov Chain Monte Carlo (MCMC) method.

An explicit and computationally efficient form of the likelihood function is an essential element for this analysis approach. For many astrophysical processes, instrumental noise is almost Gaussian and uncorrelated, resulting in simple Gaussian likelihood functions. However, there exist many problems in which the noise statistics are more complex or where other physical stochastic processes complicate the calculation of the likelihood \citep{2022arXiv220501866P}.

An alternative approach to computing the posterior distribution is through Monte Carlo simulations of the data. In this case, to compute the probability of the data given the model, instead of writing an explicit form for the likelihood function, one tries to approximate it with Monte Carlo simulations. This is done by drawing parameters from the prior distribution to produce \emph{noisy} simulations (where the noise is generated from the known noise model) and accepting models that produce a perfect match (i.e., an exact reproduction of the data). Theoretically, the distribution of the parameters of all accepted models approximates the true posterior. This approach requires a fast simulation pipeline that can produce noisy realizations of data, but it does not require an explicit form for the probability density of noise and therefore does not require an explicit likelihood.

The main challenge, however, is that the probability of generating a data set $d_{simulated}$ with a small distance to the observed data $d_{observed}$ significantly decreases as the dimensionality of the data increases. For high-dimensional data sets (e.g., an image with many pixels), the rate of replicating the exact data within a reasonable numerical precision is extremely low (for practical purposes, zero). In other words, one may need to produce an uncountable number of simulations before one can randomly reproduce an exact (or near-exact) copy of the image being analyzed.

Implicit likelihood inference refers to all methods that, by making certain approximations, facilitate this procedure. They are generally designed to 1) reduce the dimensionality of the data, in effect compressing it, and 2) accept models that are deemed \emph{sufficiently close} to the observed data (instead of only accepting perfect matches), and sometimes 3) amortize the calculations so that repeating the simulations is not needed for new inference. 

Approximate Bayesian Computation \cite[ABC,][]{10.1093/genetics/162.4.2025, 2009arXiv0904.0635B}, for example, is one such approach. Here, the dimensionality reduction is done by calculating a set of lower-dimensional summary statistics, $S$, which are selected to capture the relevant information in $d_{observed}$. Additionally, all simulations which result in a predicted statistic within a small distance $\epsilon$ from the same statistic calculated for the observed data, $S_{observed}$, are accepted. Here,  $\epsilon$ is chosen to be small enough so that the posterior is not significantly widened while allowing a reasonable number of accepted simulations.

If the chosen statistic is a sufficient statistic (meaning that it contains all of the available information relevant to the parameters being estimated), the optimal posterior can be obtained. However, if the statistic is not sufficient, precision is lost, and the posterior is widened, but the analysis remains accurate. In this case, effectively, the information lost through compression does not contribute to the posterior.

One of the drawbacks of ABC, however, is that the posterior calculation is not amortized; for every new data set, one needs to reproduce a large number of simulations from the prior and only keep the accepted ones. The method could also be somewhat sensitive to the arbitrary choice of $\epsilon$ in multi-dimensional spaces. This could result in inefficient sampling (too few accepted simulations) or loss of information (posterior smearing).

\subsection{I.L.I. and machine learning}

With the advent of powerful machine learning methods, implicit likelihood inference has seen a rapid expansion \citep[see ][]{2019arXiv191101429C}. Machine learning has been used to contribute to two different aspects of this inference framework: 1) to discover informative compressed statistics and 2) to model the distribution of the parameters of the simulated data and the compressed statistics. 

In traditional implicit likelihood methods, finding a suitable and informative statistic that can allow efficient compression of the data without significant information loss could be challenging. In many cases, such statistics are simply \emph{guessed} \citep[e.g.,][]{2017JCAP...05..037B}. Despite the fact that a bad guess does not introduce biases or inaccuracies, it does reduce the amount of retained information, resulting in a loss of precision. 

Deep learning methods have been shown to be efficient tools for data compression (e.g., with an autoencoder), making them a promising candidate for this operation. An example of this are the Information Maximising Neural Networks \cite[IMNN,][]{2018PhRvD..97h3004C}, where neural networks are trained to calculate arbitrary informative statistics by maximizing the Fisher information as the training objective. This results in statistics that are highly informative, but the procedure involves the calculation of the Fisher matrix during training, which could be complex and computationally expensive.

In this work, we show that a point estimate of the parameters provided by a deep learning model trained with a mean-squared error loss can be treated as a compressed statistic of the data, which can then be used within an implicit likelihood inference framework to obtain samples from the posterior. This means that implicit likelihood inference can quantify the uncertainties of the predictions of any neural network that has been trained to produce a point estimate of any parameter.

The second aspect of the application of machine learning to implicit likelihood inference is to model the distribution of the parameters of the simulations and the compressed statistics (the predictions) computed from the simulations and data. Instead of simple methods like ABC, these methods typically fit a model to a collection of simulations and the resulting statistics. Therefore, these models are amortized as they can be used to perform inference for many different observations without the need to generate new simulations. This makes it highly efficient when compared to methods such as ABC.

\subsection{Choice of implicit likelihood inference structure}

As described in \citet{2019arXiv191101429C}, there exist many different possibilities to structure an implicit likelihood inference framework using machine learning models. Here we explain our choice and describe the motivation for this specific setup. 

Our goal is to obtain an amortized model to infer the posterior of the macro-parameters of strong gravitational lenses in a way that is scalable to the hundreds of thousands of lenses that will be discovered by upcoming surveys. We are interested in the inference of a set of $\mathcal{O}(10)$ parameters describing the variables of simple-parametric models like the SIE and S\'{e}rsic profiles. Therefore, methods like neural ratio estimators \cite[e.g.,][]{2019ApJ...886...49B, 2020arXiv201007032C, 2019arXiv190304057H}, which are typically only practical for lower-dimensional inference problems, do not seem to be suitable here. We therefore focus on direct density estimation methods.

We train a regression convolutional neural network (hereafter referred to as CNN) to predict a point estimate of the lens parameters. In what follows, we refer to the vector of the true lens parameters as $\theta$ and the prediction of the networks as $\hat{\theta}$. Since our neural networks represent a function of the data alone with no other stochastic or fixed variables in them, we can then consider these predictions, $\hat{\theta}$, to be compressed statistics of the data. Of course, it is clear that these compressed statistics are highly informative about the parameters of interest, $\theta$, since they have been trained to predict them. This choice of compressed statistics is primarily motivated by the simplicity of the operation: there is no need to compute expensive Fisher matrices and the labels for $\theta$ provide a simple supervised method to train this regression network on an explicit target. 

Once trained, a large set of simulations $\textbf{S}=\{\mathrm{S_1},\mathrm{S_2},...,\mathrm{S_N}\}$ with parameters drawn from the prior are produced. These simulations are given to the CNN, and the combination of the true and the predicted parameters are saved to produce the training data $\textbf{P}=\{(\theta_1,\hat{\theta}_1),(\theta_2,\hat{\theta}_2), ..., (\theta_N,\hat{\theta}_N) \}$ for a density estimation model, which, once trained, can estimate the posterior, $p(\theta | \hat{\theta})$. 

To achieve this, an important decision needs to be made whether to model the likelihood, $p(\hat{\theta} | \theta)$, the posterior, $p(\theta | \hat{\theta})$, or the joint distribution, $p(\theta , \hat{\theta})$ of the variables \citep[see ][]{2019MNRAS.488.4440A}. Of course, in all cases, it is ultimately the posterior that is of interest. However, when modeling the densities, it is possible to choose any of these three options. 

In this work, we chose to model the likelihood using a mixture density network \citep{370fbeadb5584ba9ab2938431fc4f140}. Given the true parameter $\theta$, our mixture model predicts the likelihood, $p(\hat{\theta} | \theta)$. Then, at inference time, by performing an MCMC over the true parameters and including any arbitrary prior, we can obtain the posterior. Note that the MCMC is done on the mixture distribution, which is extremely light, resulting in fast convergence. 

The main advantage of this method is that it does not hard code the prior into the learned posteriors. Its flexibility allows using different priors for different lensing systems, or to update the prior and to redo the analysis with minimal overhead. For example, in strong lensing analysis, it is common to use the stellar light of lensing galaxies, their kinematics, or environments to place priors on the parameters of foreground galaxies on a case-by-case basis \citep{2019A&A...632A..36C, 2020ApJS..247...12S}. This method can also accommodate complex priors or priors with sharp features (e.g., the edges of flat distributions). Sampling the priors during the MCMC stage allows for posteriors of data with arbitrarily complex priors to be modeled accurately.  

We also train an approximate Bayesian neural network (BNN, see section \ref{bnn} of the Appendix) to compare its results with the posteriors obtained by the methods proposed in this work. To be able to have a controlled comparison between these two methods, instead of using point estimates from simple CNNs as mentioned above, we use the mean of the predictions of this BNN as the compressed statistics for our implicit likelihood inference model. This ensures that any potential differences in accuracy between the two models cannot be attributed to factors like insufficient training, different expressivity of the networks, etc. Moreover, we chose the prior distribution used in generating the BNN's training data to be the same as the one used during posterior inference with the MDN method. The motivation for this choice is to have the BNN and MDN approach both perform posterior inference with the same prior distribution (in the BNN method, the prior is implicitly learnt from the training data). Therefore, in the limit of perfect performance (and using lossless summary statistics $\hat{\theta}$ for the MDN approach), both methods should predict the same posterior distribution. We emphasise that this choice is simply made here to allow for a clear comparison and that normally the predictions of a simple feed-forward CNN would be equally suitable.

\subsection{Mixture Density Networks}
\label{density_est}
Mixture density networks (MDNs) are neural networks that model conditional probability densities as a mixture of parametric distributions. Typically, the parametric model is chosen to be a mixture of Gaussian distributions defined as

\begin{equation}
    \label{gmmeq}
    p(\theta | x, w)  = \sum_k^K \phi_k(x, w) \mathcal{N}(\theta; \mu_k(x, w), \Sigma_k(x, w)),
\end{equation}
where $\phi_k$, $\mu_k$ and $\Sigma_k$ represent the weight, mean and covariance matrix for each of the $K$ components. An MDN models a conditional probability density parameterized by a fixed set of trained network weights, $w$. 

As inputs, the network receives $\theta$ and the Gaussian noise standard deviation used in generating the lensing simulation and outputs a prediction for the parameters $\phi_k$, $\mu_k$, and $\Sigma_k$ that model the distribution over $\hat{\theta}$. Consequently, these parameters are conditioned on the input given to the network, which, in this case, gives us a model for the likelihood $p(\hat{\theta}|\theta)$. 

In order to sample from the posterior, we use the predicted MDN likelihood function with the affine-invariant MCMC algorithm implemented in the Python package PYDELFI \citep{2019MNRAS.488.4440A}. The algorithm is written entirely in Tensorflow, allowing for GPU accelerated MCMC sampling.
For a single lens this sampling takes about 120 seconds on a single GPU, which, of course, could be simply run in parallel for different lenses (an embarrassingly parallel problem). Running in parallel, we can effectively perform the inference of 50 systems in 120 seconds using a single GPU.

Our MDN is a feed-forward neural network composed of two hidden layers (in addition to the input and output layer) each containing 256 fully-connected neurons. The ELU activation function is used with a negative $\alpha$ slope of 1.0 in all layers except the final layer, where the network's predictions follow from the same activations used in the final layer of the BNN. From our experiments, we have found that the ELU activation function greatly improves the convergence rate in comparison to other activations such as $tanh$. The number of Gaussian mixture components was increased until there was no noticeable improvements in the predicted posteriors based on the accuracy tests shown in Figure \ref{cal_default} and \ref{cdf_default}. In the end, we decided to stop at six components for the Gaussian mixture model for the likelihood density $p(\hat{\theta} | \theta)$. Given that we only predict the triangular part of the covariance matrix, which contains $N(N + 1)/2$ values for $N$ dimensions, and that the lensing parameters form a 13 dimensional vector, the number of outputs (including $\mu_k$ and $\phi_k$) for the final layer is 
$13 + 13 (13 + 1)/2 + 1 = 105$ (for the mean, the covariance, and the weight of each component) times the number of mixture components $K$. For a six component Gaussian mixture model, the total number of outputs is $630$.

The MDN's training data is composed of 100 million examples of BNN predicted compressed statistics along with the associated true simulation parameters and Gaussian noise standard deviation used for each lensing example, $\textbf{P}=\{(\theta_{1},\hat{\theta}_{1},\sigma_{\mathrm{noise},1}),(\theta_2,\hat{\theta}_{2},\sigma_{\mathrm{noise},2}), ..., (\theta_{N},\hat{\theta}_{N},\sigma_{\mathrm{noise},N})\}$. This data represents a new set of simulations independently generated from the BNN training data.

Before feeding $\theta$ to the network, we fix the value of each dimension in $\theta$ to be within the range $[-1, 1]$ using the boundary values of the uniform prior distribution $p(\theta)$. During training, we found that this improves the convergence rate towards an optimal solution while also making training more stable. We train the MDN using the AdaMax variant of the Adam optimizer \citep{2014arXiv1412.6980K} with an initial learning rate of $5 \cdot 10^{-4}$ and an exponential decay schedule with decay rate of $0.8$ applied every $75$ epochs. From our experiments, we found that using AdaMax instead of Adam further increased stability during training. We train the MDN for 1250 epochs with the batch size set to 20000. Similar to a BNN, the MDN is trained to minimize the negative log-probability of the predicted distribution evaluated at $\hat{\theta}$.

\subsection{Simulations}

Strong lensing is an image distortion that preserves surface brightness. A point in the background source at true location $\beta$ is observed at a point in the observed image plane $\theta$ following

\begin{equation}
    \beta = \theta - \alpha.
\end{equation}
where the relation depends on the scaled deflection angle $\alpha$ computed from the mass distribution in the lensing structure.

Our simulations consist of a Singular Isothermal Ellipsoid (SIE) foreground galaxy with added external shear and an elliptical S\'{e}rsic background source. As in previous works \cite[e.g.,][]{2017Natur.548..555H, 2017ApJ...850L...7P, 2022arXiv220611279S, 2022arXiv220710124S,Legin:21,Legin22}, we use cartesian coordinates $(\epsilon_x, \epsilon_y)$ for the ellipticity of both foreground galaxy and background source. This is a more suitable choice of parameters compared to ellipticity and orientation angle since for more circular morphologies the orientation angle becomes effectively unconstrained. In total, seven parameters are needed to describe the lens and six to describe the background source.

\begin{table}[ht]
    \begin{tabular}{ll}
    \toprule
    \textbf{Component}                                                    & \textbf{Distribution}                       \\ \midrule
    \textbf{Lens: SIE}                                                    &                                             \\
    Einstein Radius                                                  & $\theta_E \sim\mathcal{U}(0.8, 2)$          \\
    x-direction ellipticity                                               & $\epsilon_{xl} \sim \mathcal{U}(0, 0.6)$    \\
    y-direction ellipticity                                               & $\epsilon_{yl}\sim \mathcal{U}(-0.6, 0.6)$  \\
    x-coordinate center                                                   & $x_{lens}\sim \mathcal{U}(-0.4, 0.4)$       \\
    y-coordinate center                                                   & $y_{lens}\sim \mathcal{U}(-0.4, 0.4)$       \\ \midrule
    \textbf{External shear}                                               &                                             \\
    x shear component                                                     & $\gamma_1 \sim \mathcal{U}(-0.3, 0.3)$      \\
    y shear component                                                     & $\gamma_2 \sim \mathcal{U}(-0.3, 0.3)$      \\ \midrule
    \textbf{Source: S\'{e}rsic ellipse} &                                             \\
    Effective radius                                                      & $R_{eff} \sim \mathcal{U}(0.1, 0.2)$        \\
    x-direction ellipticity                                               & $\epsilon_{xs} \sim \mathcal{U}(0, 0.4)$    \\
    y-direction ellipticity                                               & $\epsilon_{ys} \sim \mathcal{U}(-0.4, 0.4)$ \\
    x-coordinate center                                                   & $x_{src} \sim \mathcal{U}(-0.3, 0.3)$       \\
    y-coordinate center                                                   & $y_{src} \sim \mathcal{U}(-0.3, 0.3)$       \\
    S\'{e}rsic index                    & $n \sim \mathcal{U}(1, 4)$    \\ \bottomrule         
    \end{tabular}
    \caption{Prior distribution on all lensing parameters. Each parameter is sampled from a uniform distribution.}
    \label{tabprior}
\end{table}

\begin{figure}
    \centering
    \includegraphics[width=1.0\linewidth]{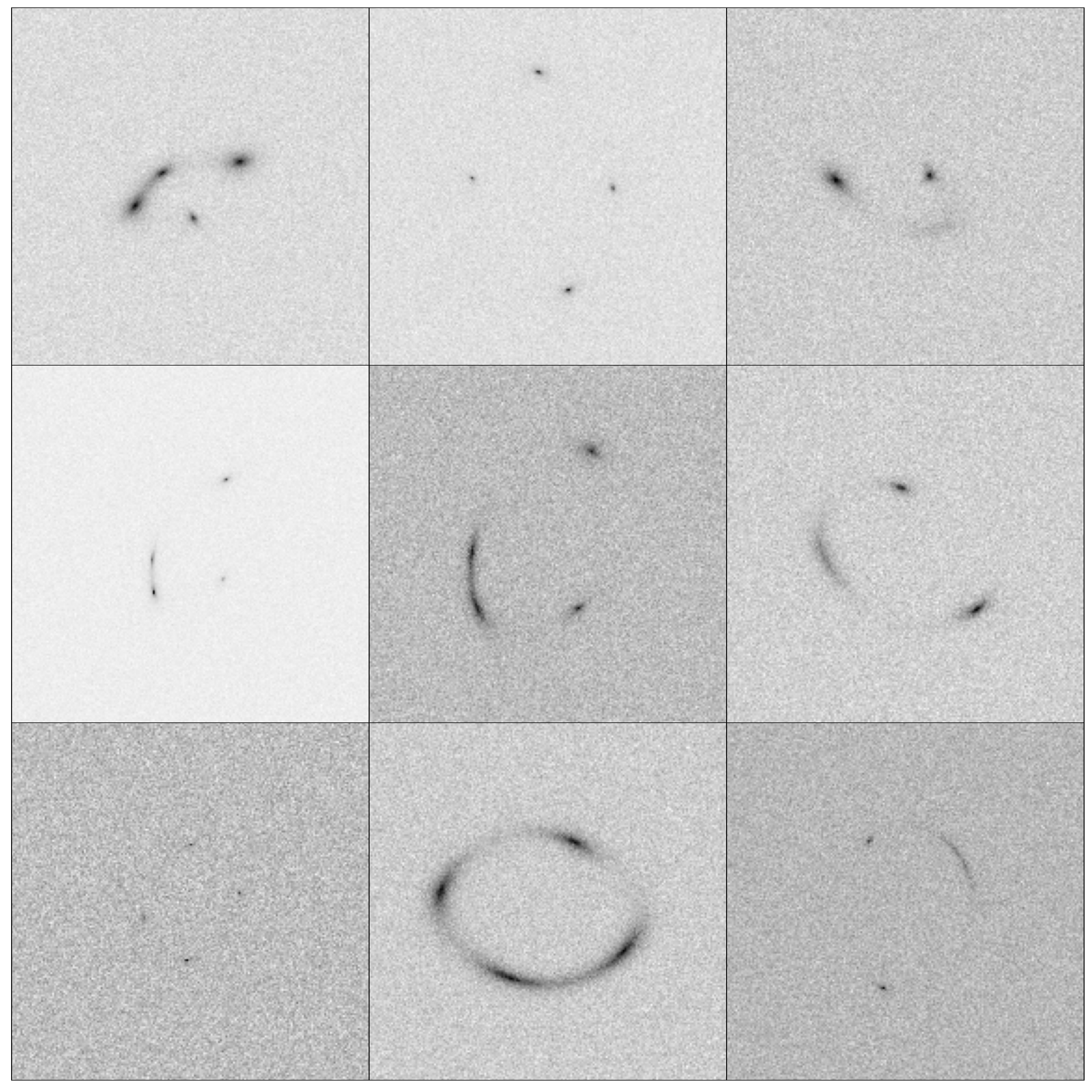}
    \caption{Examples of simulated strong gravitational lensing images. The simulations span a wide range of possible configurations of strong lensing systems consisting of a SIE primary lens with external shear and an elliptical S\'{e}rsic background source.}
    \label{simulations}
\end{figure}

The source plane spans a field of view of 3 arcseconds while the lens plane spans a width of  7.68 arcseconds. Both planes have a spatial resolution of 192 by 192 pixels. We generate simulations by randomly sampling the lens and source parameters from a uniform distribution, spanning a wide range of lensing configurations. This represents our prior distribution on the lensing parameters $p(\theta)$, and details regarding its range are found in table \ref{tabprior}. Each simulation contains pixel-wise independent Gaussian noise with standard deviation chosen from a uniform distribution between 1\% and 10\% of the peak lens surface brightness. Examples of our lensing simulations are shown in Figure \ref{simulations}. 

The simulation code is written entirely in Tensorflow, resulting in fast GPU-accelerated generation of strong lensing images which will be used for on-the-fly generation of training data for the Bayesian neural networks and predictions of compressed statistics.

\section{Tests and Results}
\label{results_tests}

\begin{figure}
    \centering
    \includegraphics[width=1.0\linewidth]{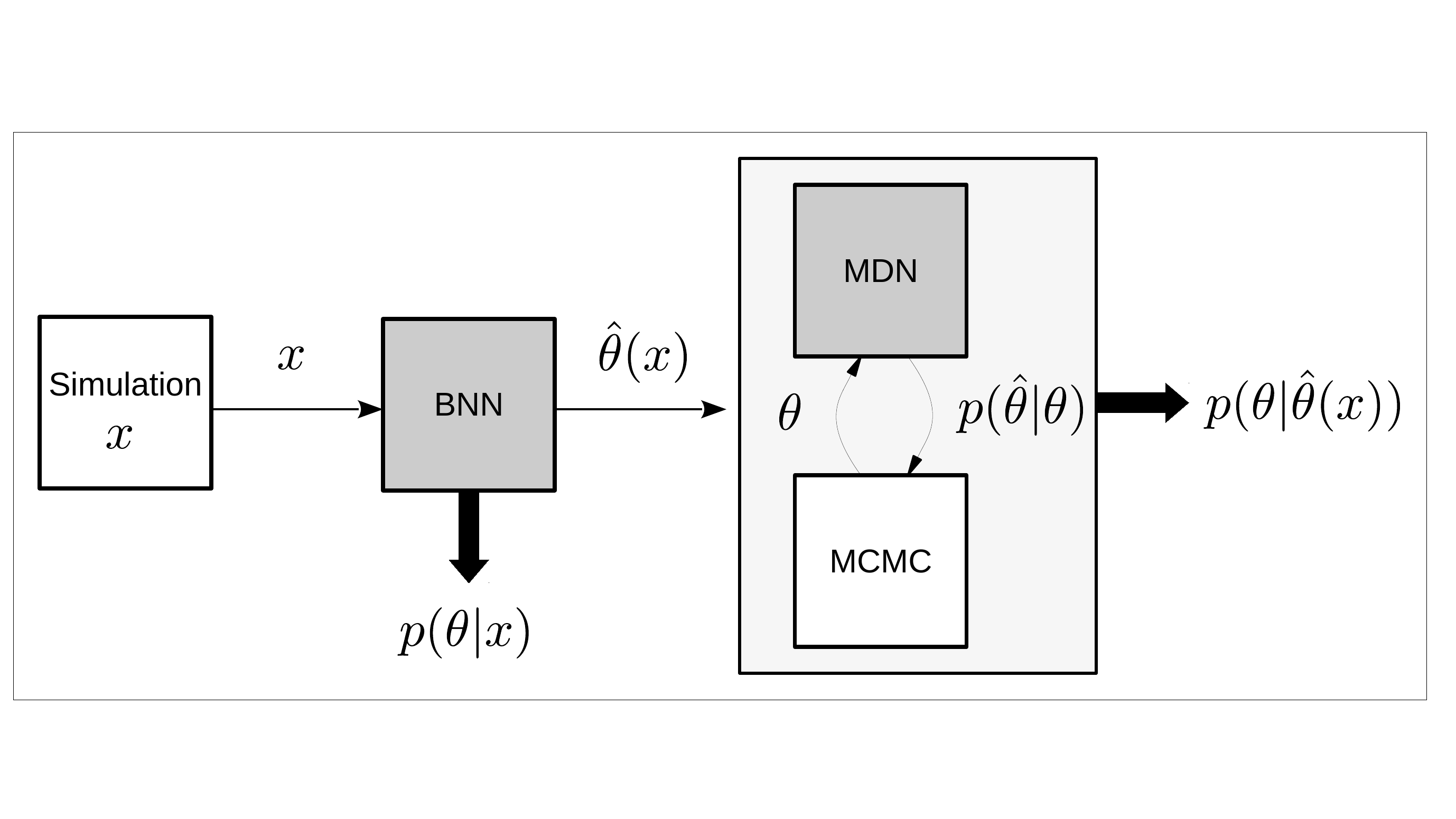}
  \caption{Complete setup of the inference problem. Lensing simulations $x$ with added observational noise are generated from randomly sampling model parameters from the prior distribution and fed to the BNN for compression into predicted lensing parameters. These compressed statistics are treated as low-dimensional observations of the data $x$ for which the modeled MDN likelihood is evaluated at when sampling the posterior $p(\theta | \hat{\theta})$ using an MCMC.}
  \label{setup_fig}
\end{figure}

\begin{figure*}
  \centering
  \includegraphics[width=1.0\linewidth]{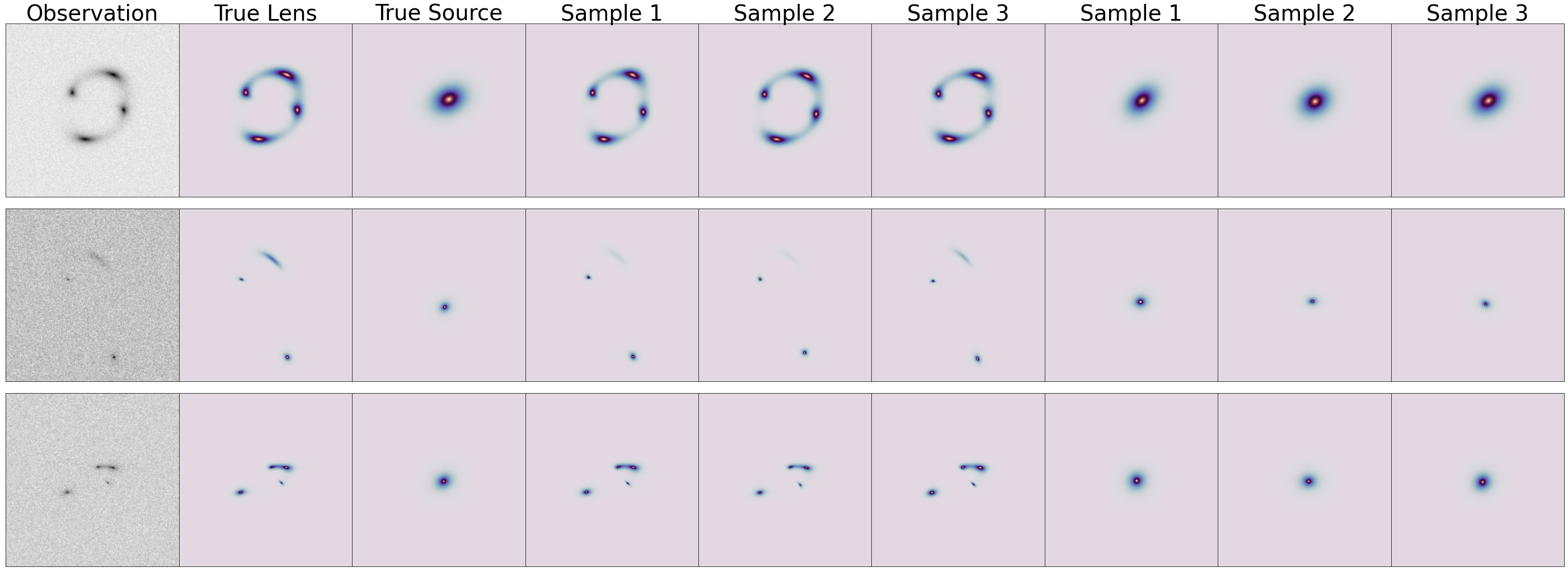}
  \caption{Examples of observations with the true lens and source model alongside simulations generated using samples of strong lensing parameters $\theta$ from predicted lensing posteriors using the MDN likelihood model. The samples from the predicted MDN-based posterior are consistent with the ground truth lensing image and background source.}
  \label{example_predictions}
\end{figure*}

We train two different BNN models: one with a Dropout rate of $10\%$ and another with a Dropout rate of $0\%$, effectively predicting a posterior distribution parametrized by a fix set of network weights. We apply both BNN models on a test set of 100000 lensing simulations in order to get  samples from their predicted posteriors $p(\theta | x)$. To do so, for every lensing simulation, the BNN performs 30 forward passes each sampling 100 times from the output distribution, giving a total of 3000 samples from $p(\theta | x)$. In the case of the $0\%$ BNN, 3000 samples are obtained from a single forward pass as the output distribution remains fix over repeated forward passes. We also get compressed statistics $\hat{\theta}$ by computing the mean of the samples from each predicted posterior. Afterwards, we sample from the alternate posterior $p(\theta | \hat{\theta})$ using the MDN likelihood and the PYDELFI Tensorflow implementation of the affine-invariant MCMC. An illustration of the setup is shown in Figure \ref{setup_fig}. Examples of samples from the predicted MDN-based posterior distribution compared to the true strong lensing images is shown in Figure \ref{example_predictions}. 

In the following section, we describe two different tests to assess the accuracy of the predicted posteriors obtained using the BNN and MDN-based models.

\subsection{Accuracy tests}
\label{acc_test}
\subsubsection{Coverage probability test}
The first test consists of calculating the coverage probabilities of the predicted posteriors following \cite{2017ApJ...850L...7P} and \cite{ 2021ApJ...909..187W}. Simply put, the test is based on the idea that for accurate posteriors, the true value should be found $x\%$ of the time in a region that encompasses $x\%$ of the total posterior probability.

For each simulation, we compute the distance between the true model parameter $\theta$ and the mean of the posterior samples $\hat{\theta}$ based on a chosen distance metric. Similar to \cite{2021ApJ...909..187W}, we define this distance metric as

\begin{equation}
    d(x) = (x - \hat{\theta}) \cdot \Sigma_{d} \cdot (x - \hat{\theta})^T,
\end{equation}
where $\Sigma_{d}$ is the empirical covariance matrix of true lensing parameters over our entire set of test simulations. Then, for each predicted posterior, we compute the fraction of samples that fall within the distance separating $\theta$ and $\hat{\theta}$. This gives us an estimate of the probability volume needed to include the true value. Afterward, we check over the entire set of simulations the number of times the truth is found within some arbitrary probability volume $x\%$. If the truth falls within this volume $x\%$ of the time, then, on average, the predicted posteriors have perfect coverage probabilities.

\begin{figure}
  \centering
  \includegraphics[width=1.0\linewidth]{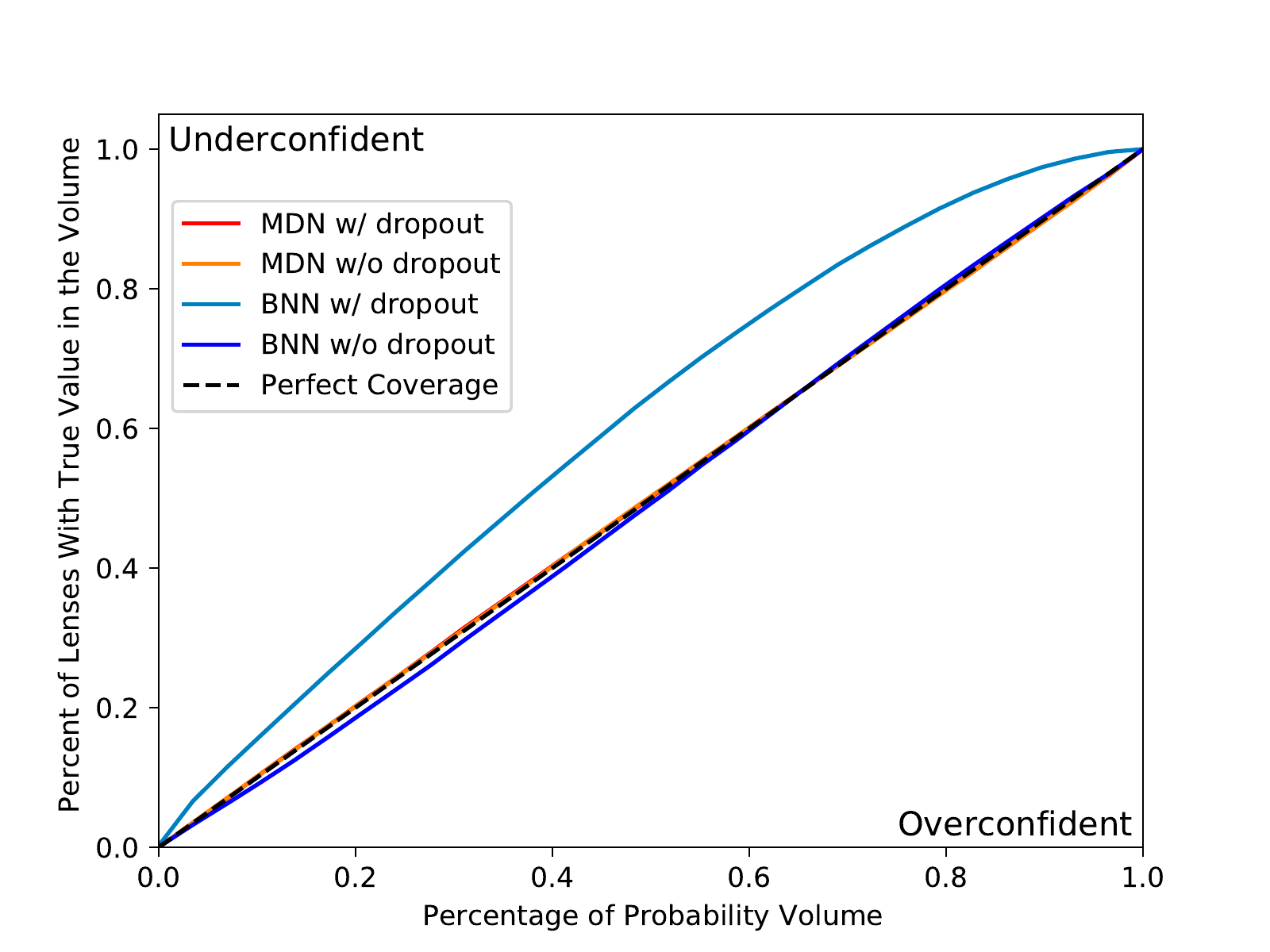}
  \caption{Results from the coverage accuracy test detailed in section \ref{acc_test} from predicted posteriors over a set of 100000 lensing simulations. A first MDN model (red) is trained to predict the likelihood $p(\hat{\theta}|\theta)$ using summary statistics $\hat{\theta}$ from the BNN with dropout (teal), and a second MDN model (orange) is trained with summary statistics from the BNN without dropout (blue). The reason for including the BNN model with dropout, which is purposely not well calibrated, is to demonstrate that the MDN model is capable of correcting for the BNN's  inaccuracy, as can be seen in this figure. This is in agreement with the statement that even if the compressor (in our case, a BNN) is poorly optimized, the MDN can correct for it by providing accurate posterior distributions given the compressor's predictions.}
  \label{cal_default}
\end{figure}

Figure \ref{cal_default} shows the results of this test on BNN-based and MDN-based sampling of the lensing posterior. In the limit of perfect coverage probabilities, the curve obtained should lie exactly on the black diagonal line. The results show that the BNN model with $10\%$ Dropout is much more conservative in its uncertainties compared to the other models. In contrast, both MDN-based methods appear to follow the diagonal line quite closely. The BNN with $0\%$ Dropout also appears to do well, although it exhibits slight overconfidence for low percentages of probability volume. This test alone shows that both MDN models can perform accurate inference of lensing posteriors, whether or not the compressed statistics have added noise due to Dropout.

\subsubsection{HPD mass distribution test}
\label{zeta_test}
The second test is taken from \cite{2015MNRAS.451.2610H} and revolves around obtaining the total probability mass contained in the highest probability density region (HPD) of the predicted posteriors. Generally, given a data point $x$ sampled from a distribution $f(x)$, we can compute this total probability mass $\zeta(x)$ as 

\begin{equation}
    \label{zeta}
    \zeta(x) = \int_{f(u) \geq f(x)} f(u) d^n u
\end{equation}
where the region of integration is defined by the constraint $f(u) \geq f(x)$. This region is known as the highest probability density region (HPD), and has the property that any data point sampled outside of it will always have lower probability than samples within. The key idea behind computing $\zeta$ is that under the null hypothesis where the data $x$ is sampled from the distribution $f(x)$, $\zeta$ is uniformly distributed. A proof of this can be found in \cite{2015MNRAS.451.2610H}.

This test can be used to verify if the true simulation parameters $\theta$ can be sampled accurately from the predicted posteriors. Given a lensing simulation, the predicted posterior now represents the reference distribution $f(x)$, and the true simulation parameters $\theta$ represent the data point $x$ defining the HPD region. Based on equation \ref{zeta}, the expression used to compute $\zeta(\theta)$ is now 

\begin{equation}
    \label{zetapost}
    \zeta(\theta) = \int_{p(u | \hat{\theta}) \geq p(\theta | \hat{\theta})} p(u | \hat{\theta}) d^n u.
\end{equation}

Computing the integral in equation \ref{zetapost} requires an explicit expression for the posterior $p(\theta | \hat{\theta})$. When performing density estimation, since we are only modelling the likelihood as a parametric distribution, we do not have access to an explicit model for the posterior. However, if sampling from $p(\theta | \hat{\theta})$ and evaluating the probability value at any sample is possible, a simple approximation to equation \ref{zetapost} is to compute the fraction of samples $u$ satisfying the constraint $p(u | \hat{\theta}) \geq p(\theta | \hat{\theta})$. Evaluating the posterior probability of a sample $\theta$ given an observation $\hat{\theta}$ can be done via Bayes' theorem

\begin{equation}
    p(\theta | \hat{\theta}) = \frac{p(\hat{\theta} | \theta) p(\theta)}{p(\hat{\theta})}.
\end{equation}

The only quantity that we do not have a closed-form expression for is the marginalized probability $p(\hat{\theta})$. Although it may be possible to model, since it is constant given an observation $\hat{\theta}$, the fraction of samples satisfying $p(u | \hat{\theta}) \geq p(\theta | \hat{\theta})$ is left unchanged, as both $p(u | \hat{\theta})$ and  $p(\theta | \hat{\theta})$ will be scaled by the same amount  $p(\hat{\theta})$. Therefore, computing the marginal probability $p(\hat{\theta})$ is not required when approximating $\zeta$.

Since it is not possible to exactly evaluate the posterior probability from a $10\%$ BNN, it must unfortunately be left out for this test. However, the $0\%$ BNN model is suitable as it predicts an explicit expression for the posterior distribution (the parametric output distribution is fixed as no Dropout is used). 

\begin{figure*}
\gridline{\fig{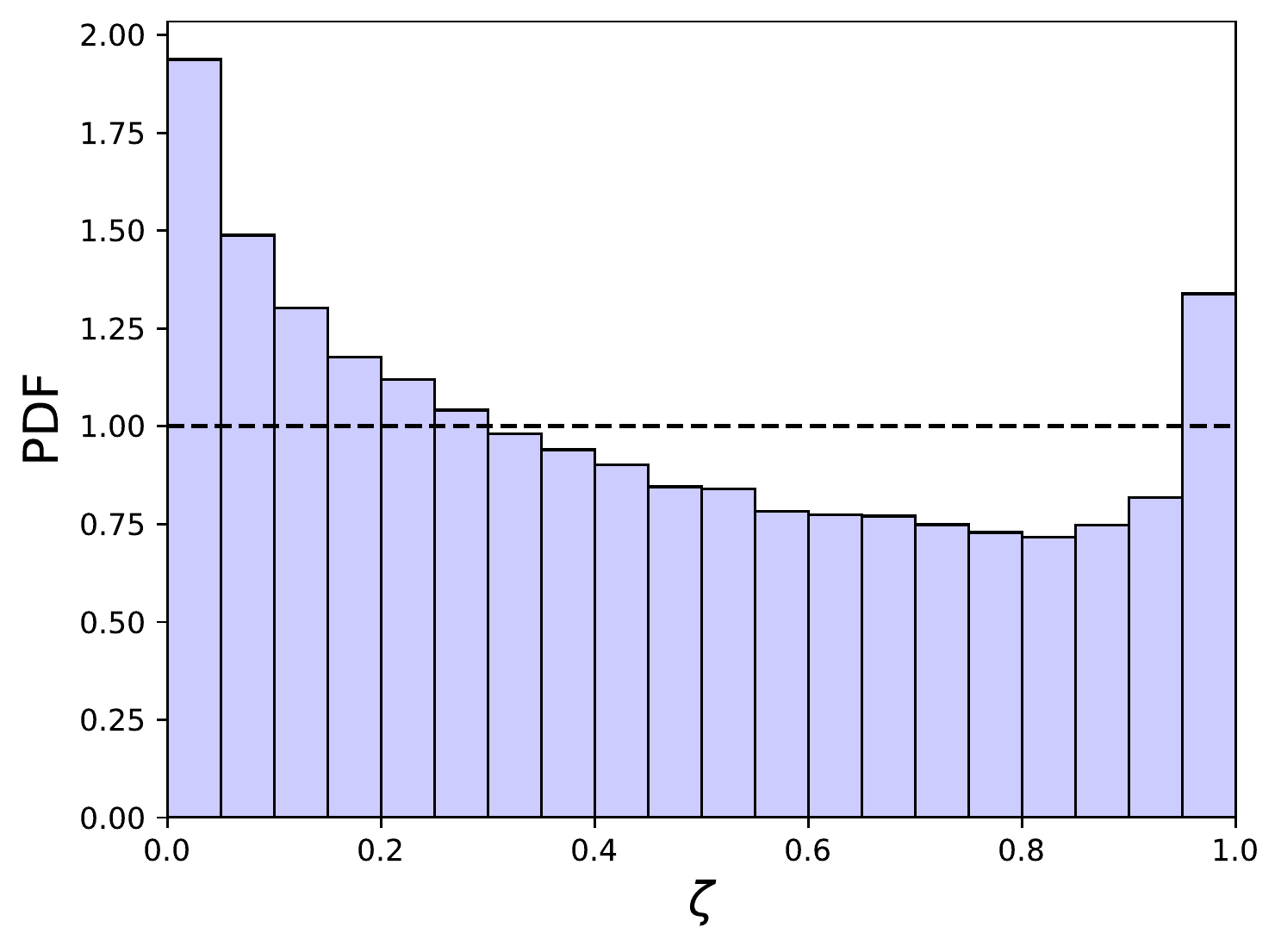}{0.33\textwidth}{(a)}
          \fig{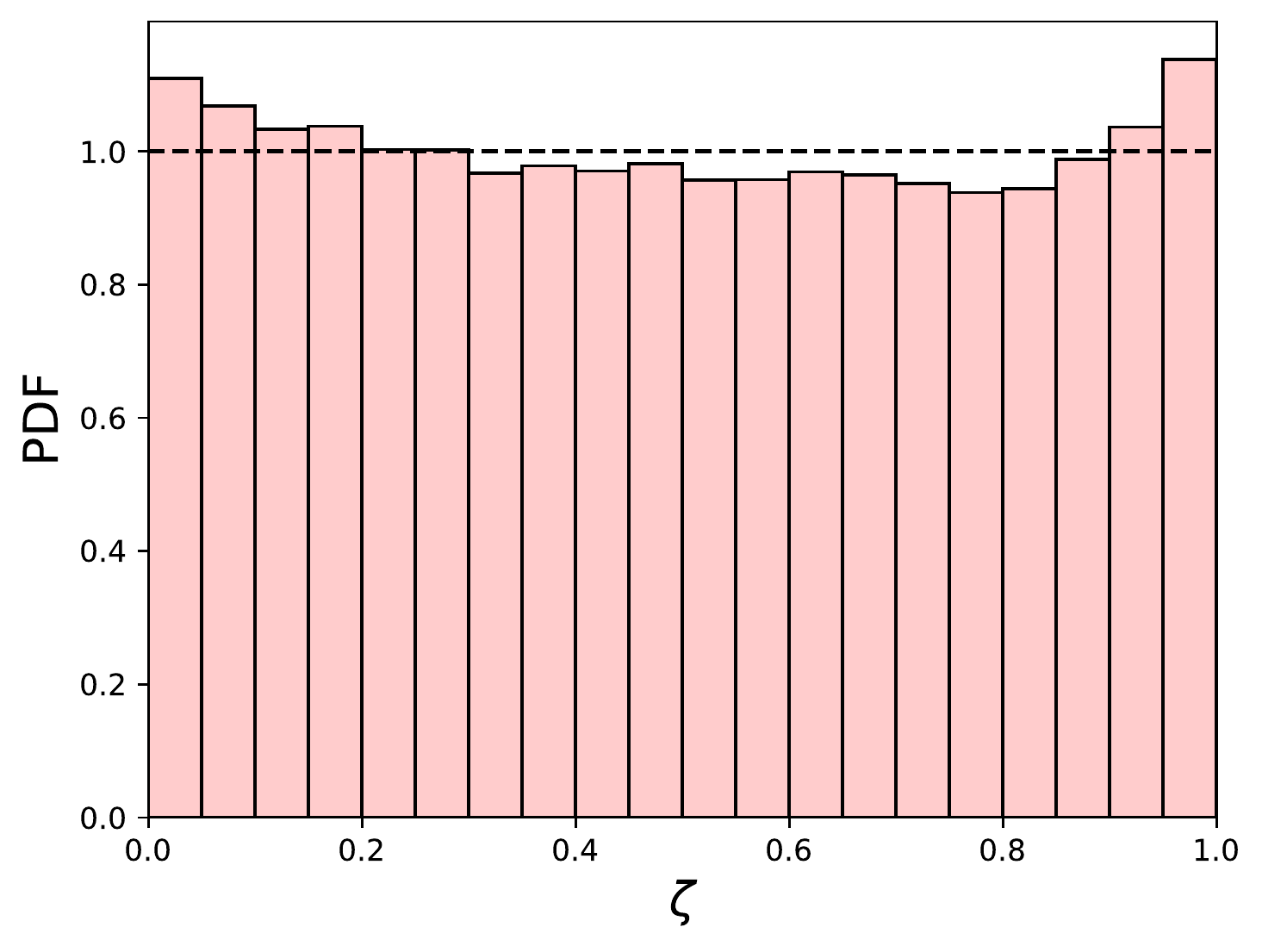}{0.33\textwidth}{(b)}
          \fig{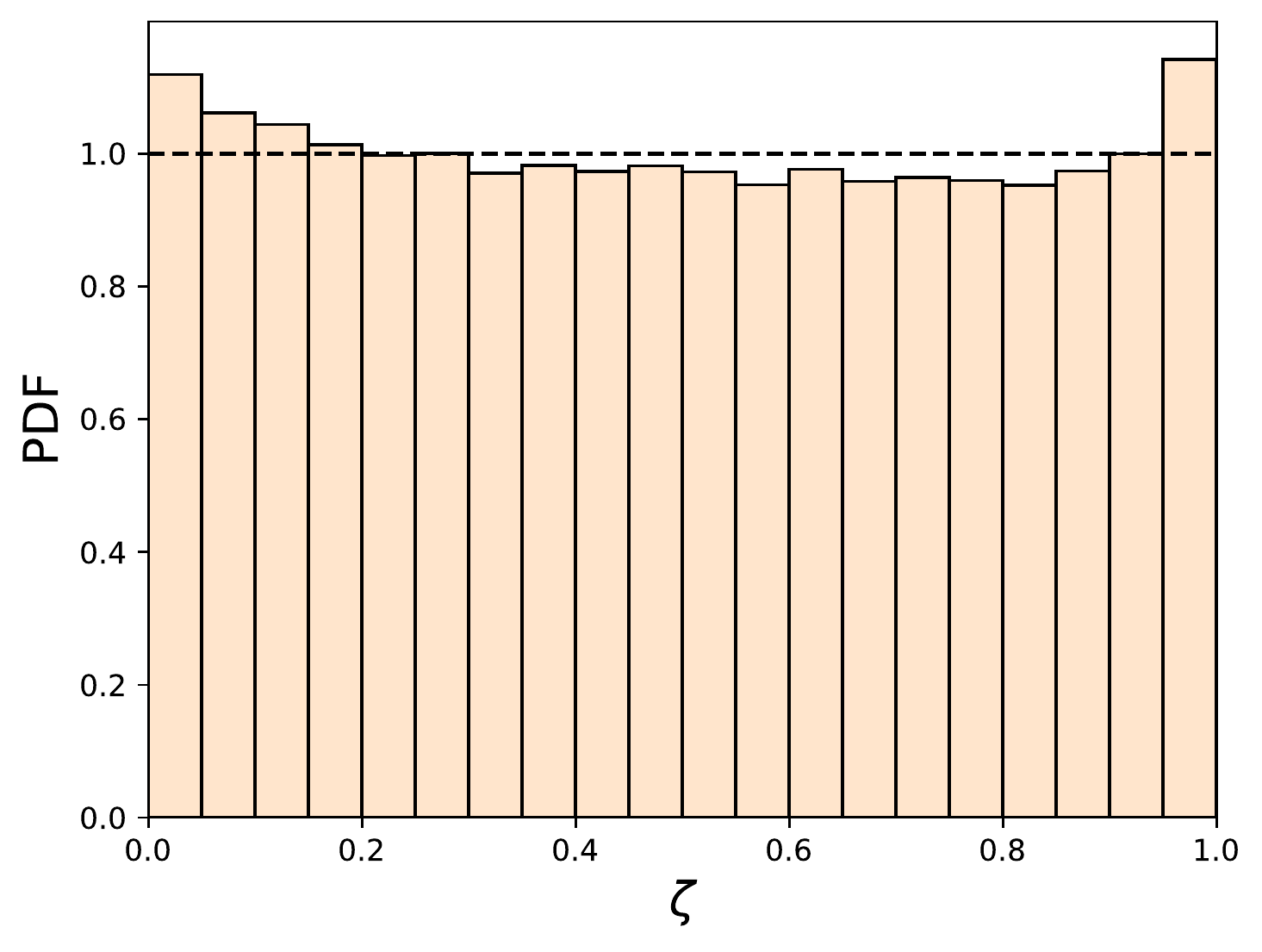}{0.33\textwidth}{(c)}}
\caption{Results of the $\zeta$ distribution based on the predicted posteriors of (a) an approximate BNN with $0\%$ dropout rate, (b) sampling MDN likelihood trained on compressed statistics from BNN with $10\%$ dropout rate and (c) from BNN with $0\%$ dropout rate. The distribution obtained using a BNN clearly deviates from that of a uniform distribution. In contrast, Figures (b) and (c) follow closely the target pdf shown as the horizontal dotted line.}
\label{hist_default}
\end{figure*}

\begin{figure}
  \centering

  \includegraphics[width=1.0\linewidth]{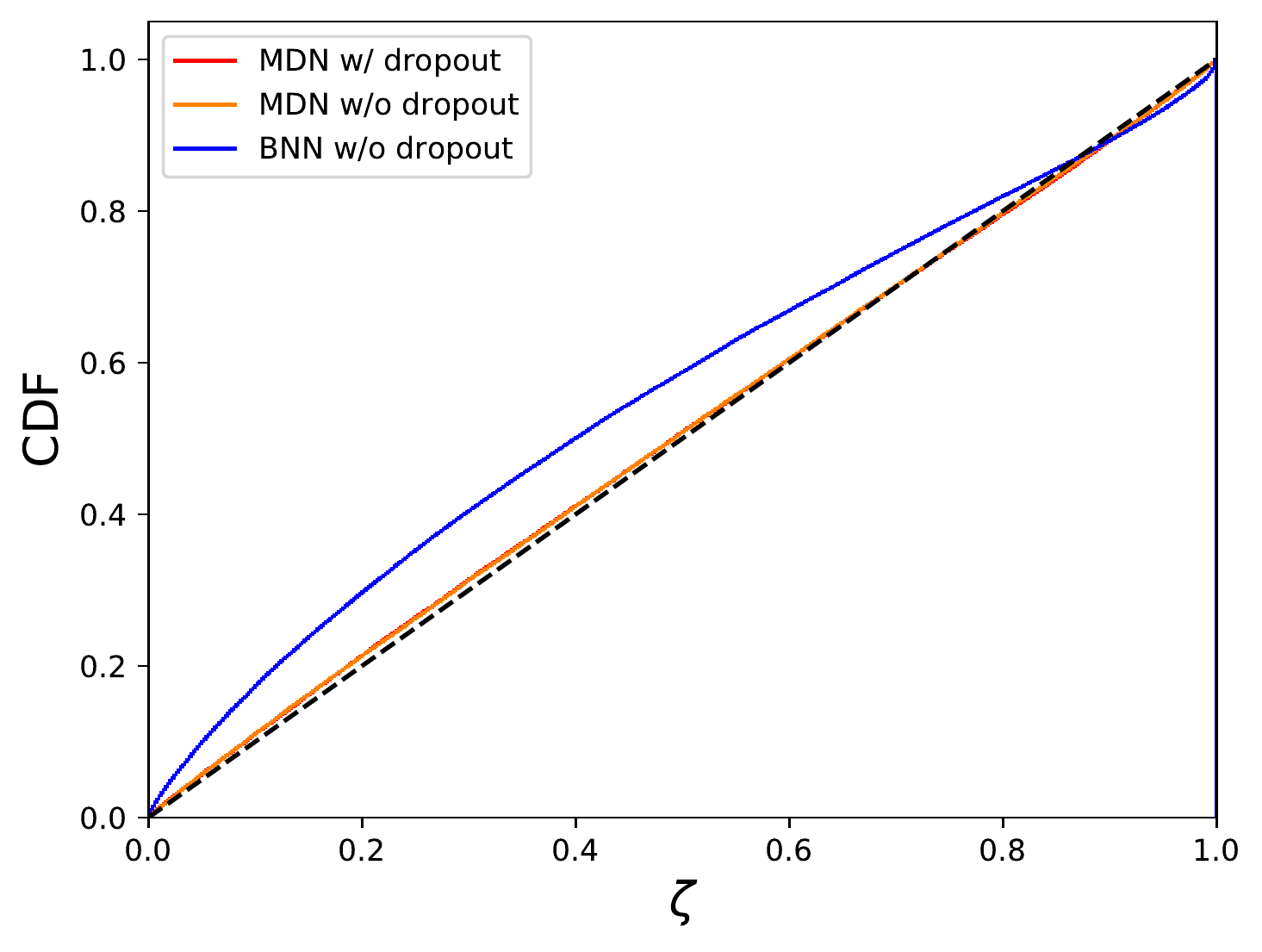}
  \caption{Cumulative distribution function of $\zeta$ distributions from Figure  \ref{hist_default}. Results from the MDN models show that they can perform posterior inference far more accurately than the BNN models.}
  \label{cdf_default}
\end{figure}

We compute $\zeta$ for every predicted posterior and associated true model parameter $\theta$ obtained from the test set of simulations. We plot a histogram of the distribution of $\zeta$ values in Figure \ref{hist_default} and its cumulative distribution function in Figure \ref{cdf_default}. The black dashed lines in Figure \ref{hist_default} and \ref{cdf_default} represent the target uniform distribution. In the $0\%$ BNN case, we clearly see the $\zeta$ distribution deviating from that of a uniform distribution, most noticeably with the presence of high amplitude bins at both extremes of $\zeta$ values, leading to a large gap between its CDF and the target CDF. While the MDN models show significant improvements towards achieving a uniform distribution for $\zeta$, they slightly exhibit similar high amplitude bins to the $0\%$ BNN model.

A concern regarding obtaining a $\zeta$ distribution from randomly generated simulations is that the posterior distribution used as the reference distribution in equation \ref{zetapost} does not remain the same for different simulations. However, this concern is resolved by the fact that $\zeta$ should be uniformly distributed independently of the reference function $f(x)$, or in our case, the predicted posterior $p(\theta | \hat{\theta})$. Therefore, the collection of $\zeta$ values obtained from different simulations should again be uniformly distributed. This is true whether the simulations are simply different realizations based on the same model parameters, or if they are generated from randomly sampled $\theta$ parameters.

\subsection{Bimodal distributions}
When performing density estimation, it is interesting to inspect the capability of the model to predict multimodal distributions present in the data distribution. For our problem, a possible bimodality is the values of the ellipticity components of the main lens galaxy $(\epsilon_{xl}, \epsilon_{yl})$ and the background elliptical source $(\epsilon_{xs}, \epsilon_{ys})$. When $\epsilon_x$ is zero, changing the overall sign of $\epsilon_y$ does not change the simulated lensing image. For noisy observations, this degeneracy could in practice extend to small but non-zero values of $\epsilon_x$, where the lensing images generated with a fixed value of $\epsilon_y$ but different signs produce almost identical data. In this scenario, the compressor network could thus predict the wrong sign for $\epsilon_y$.

\begin{figure}[ht]
  \centering
  \includegraphics[width=1.0\linewidth]{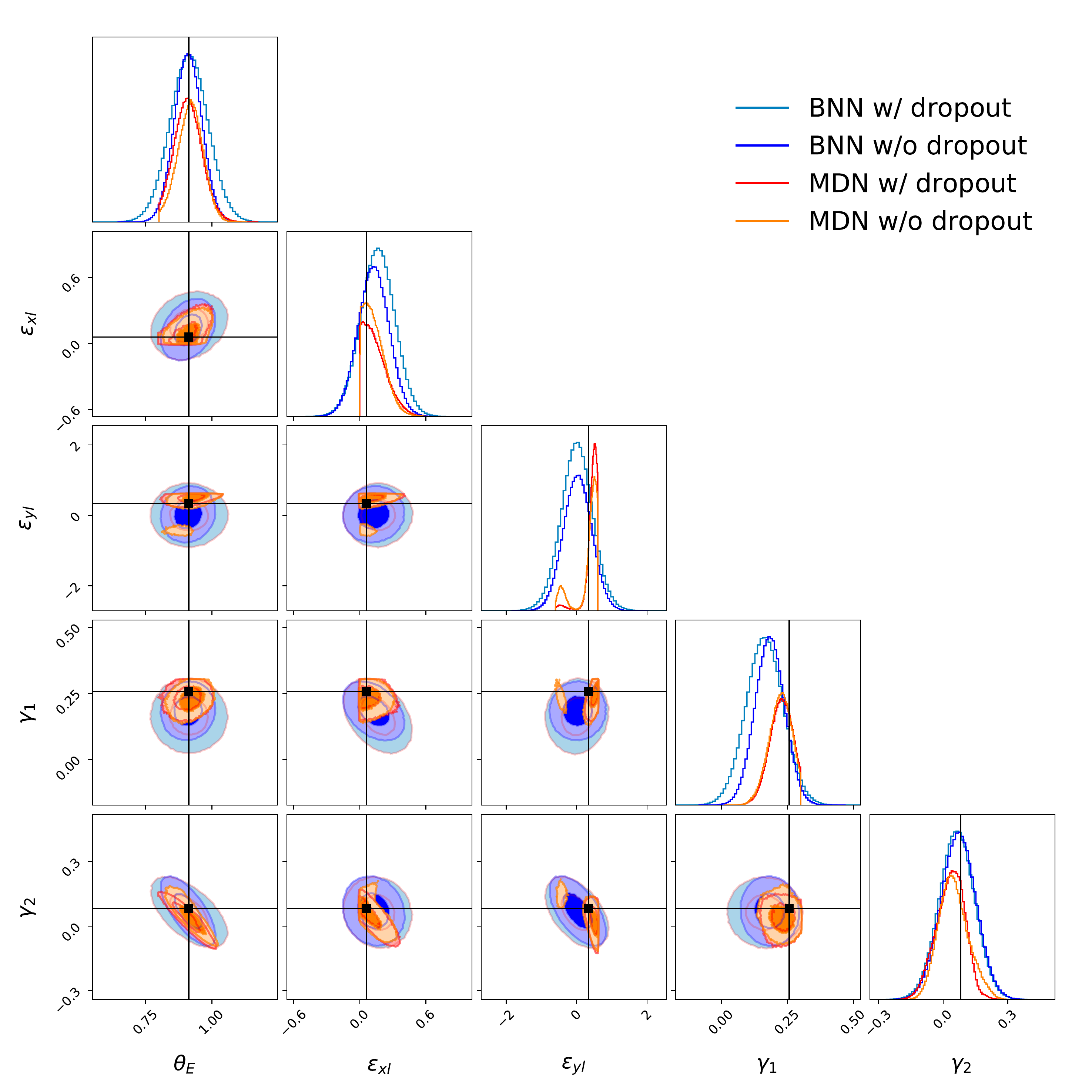}
  \caption{Example of posterior where the distribution along the $\epsilon_y$ dimension is bimodal. The peaks of the bimodal distribution appear to be centered at the same $\epsilon_y$ value with opposite signs. Both MDN-based posteriors have this bimodality.}
  \label{degen_post}
\end{figure}

By looking at the posteriors of several experiments, we did notice that our MDN models were capable of capturing this bimodality. As an example, this is shown for the marginalized posterior distribution of $\epsilon_{yl}$ from Figure \ref{degen_post}. Although most of the probability mass is centered around the true value of $\epsilon_{yl}$, a second peak is predicted for both MDN-based posteriors. Moreover, the example that we chose here also predicts a bimodal distribution for the background source's $y$ ellipticity component, $\epsilon_{ys}$, that is seen from the full posterior distribution shown in Figure \ref{degen_post_full}. In the case of the $\epsilon_{ys}$ posterior distribution, the two modes have similar probability density amplitudes.

\begin{figure}[ht]
  \centering
  \includegraphics[width=0.95\linewidth]{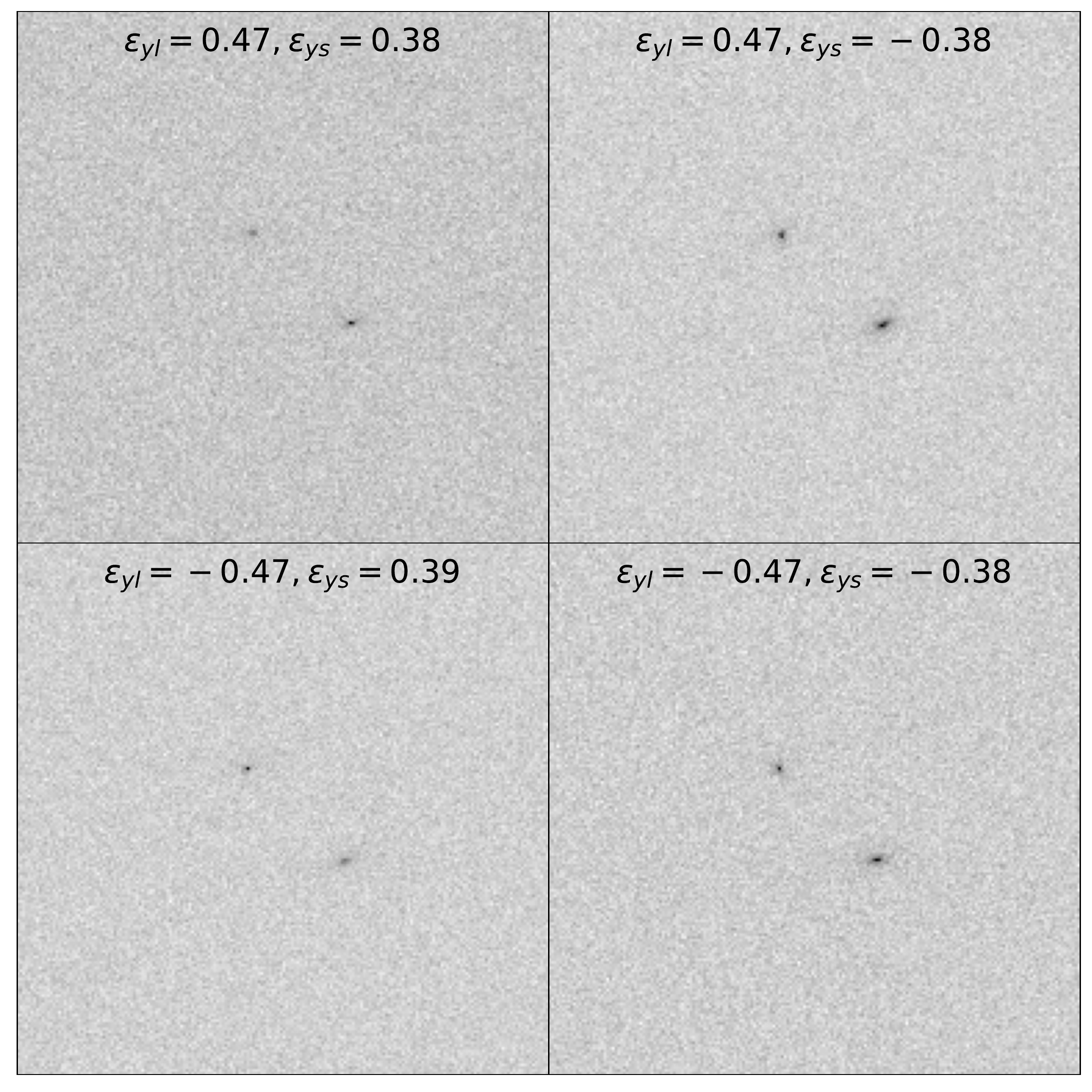}
  \caption{Example of lensing simulations generated with four samples from the predicted posterior in Figure \ref{degen_post} with different combinations of signs of both $\epsilon_{yl}$ and $\epsilon_{ys}$. The samples were picked from the MDN posterior trained on $0\%$ BNN summaries. A similar plot can be seen for samples from the $10\%$ BNN-based MDN posterior in Figure \ref{mdn0_degens}.}
  \label{degen_img}
\end{figure}

In order to verify the validity of these bimodal distributions, we generated four simulations with different combinations of signs for $\epsilon_{yl}$ and $\epsilon_{ys}$ but with values close in magnitude to their highest bimodal peak and plotted the results in Figure \ref{degen_img}. The four chosen samples were picked from the sampled posterior distribution shown in Figure \ref{degen_post_full}.

It is clear from Figure \ref{degen_img} that the lensing images share strong similarities and that the compressor network could reasonably mistake the sign of both $\epsilon_{yl}$ and $\epsilon_{ys}$. This shows that the MDN density model is sensitive enough in capturing multi-modal distributions of strong lensing parameters.

\subsection{Additional experiments}

\begin{figure*}
\gridline{\fig{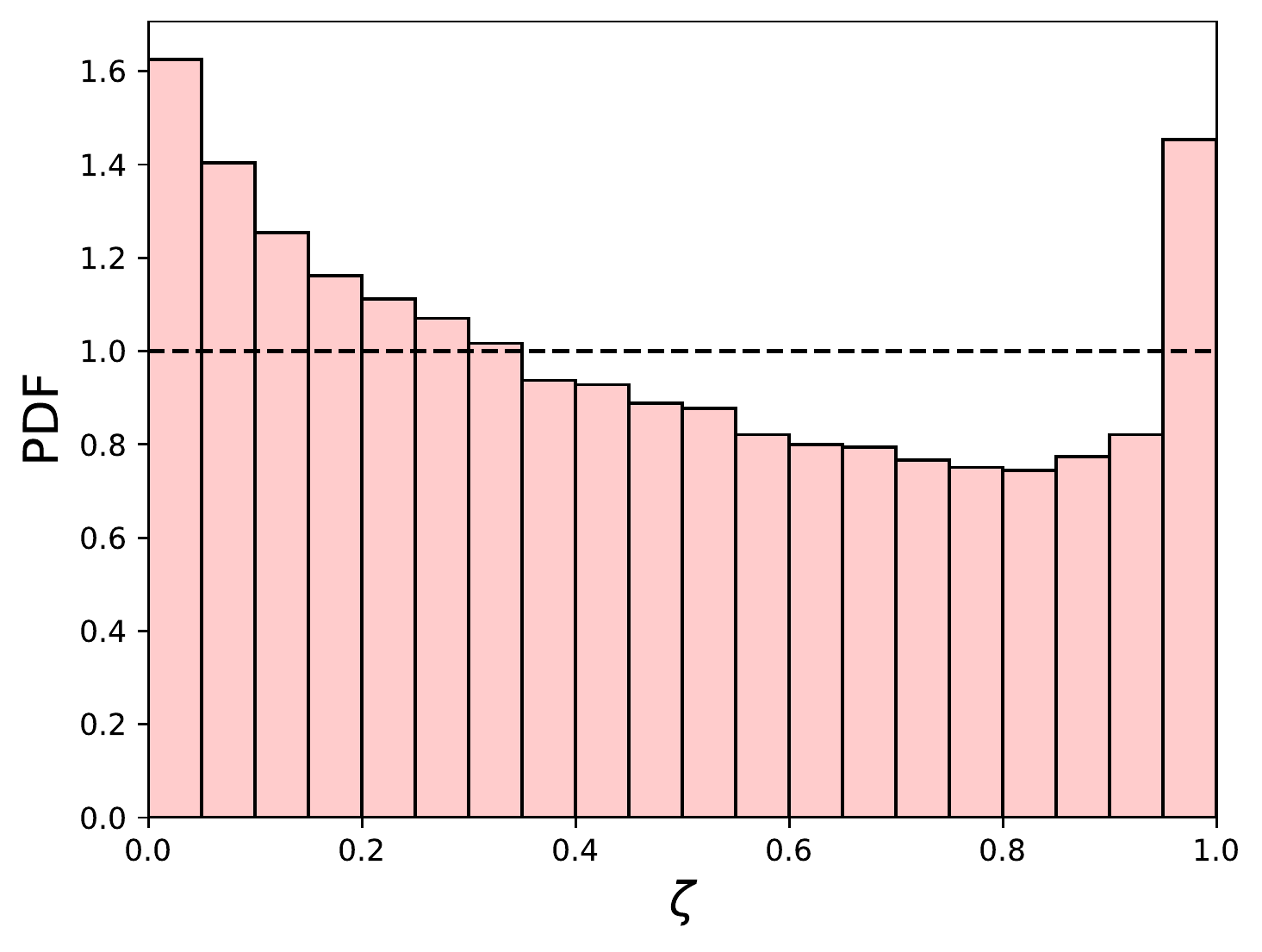}{0.33\textwidth}{(a)}
          \fig{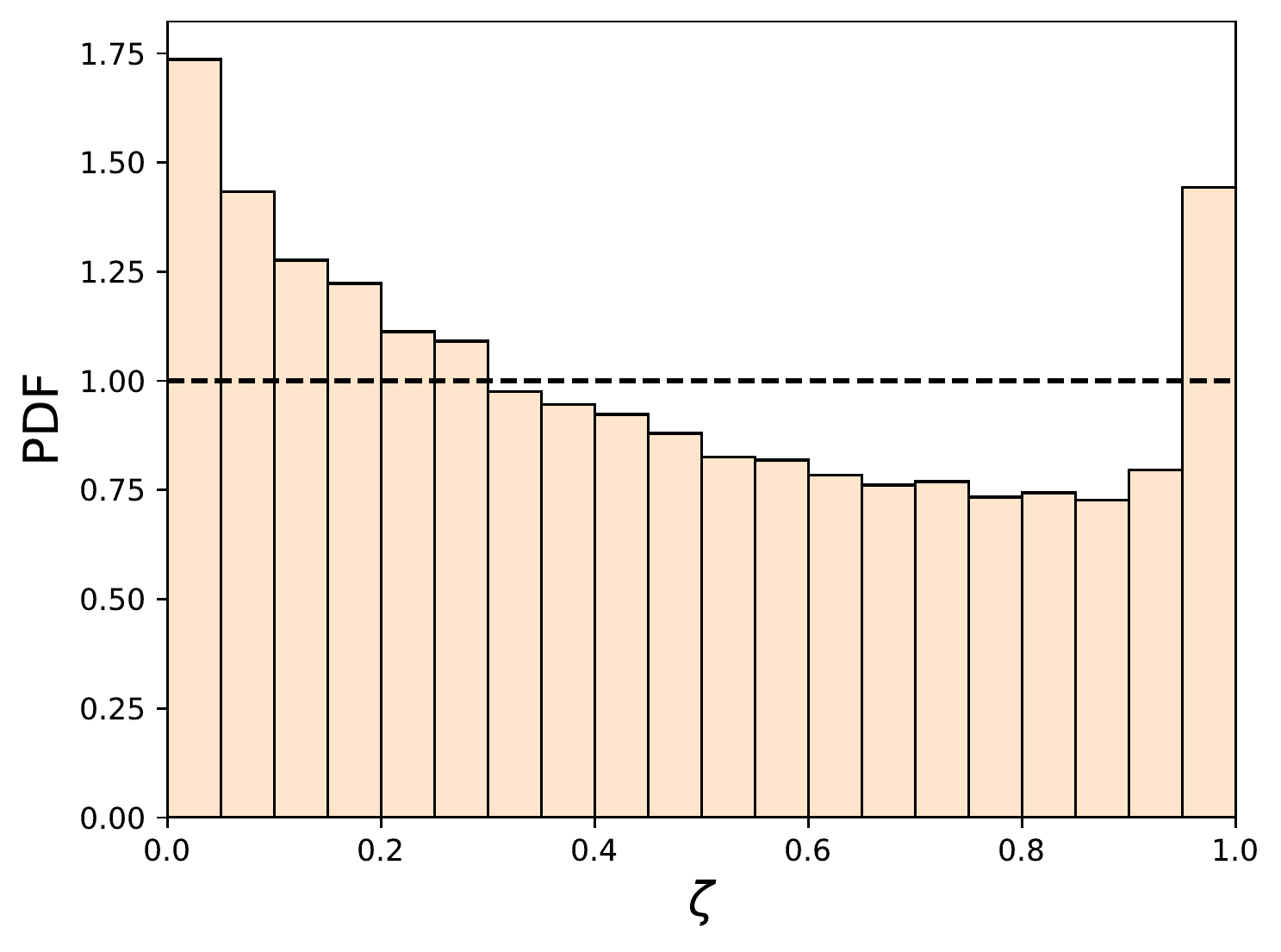}{0.33\textwidth}{(b)}
          \fig{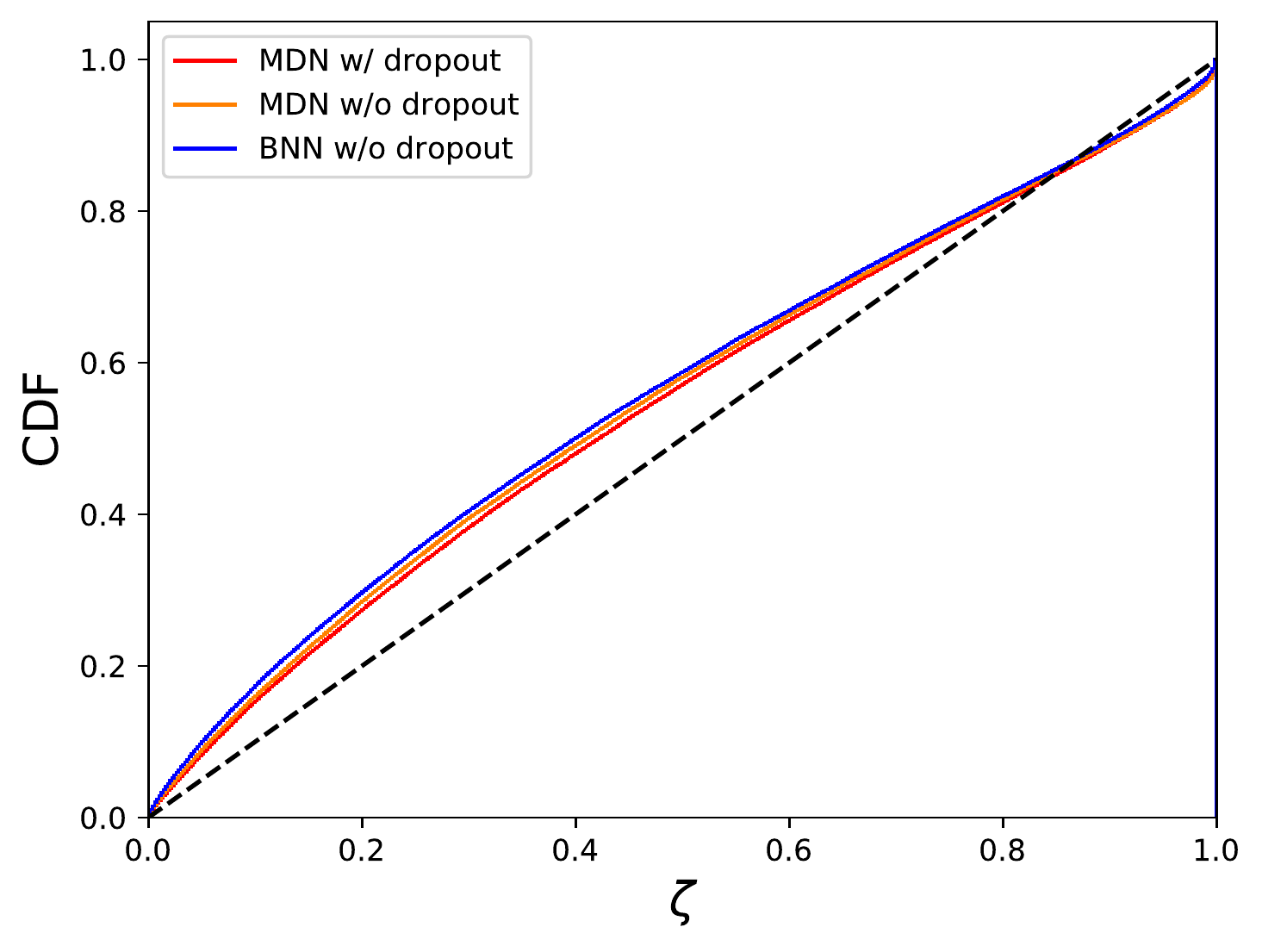}{0.33\textwidth}{(c)}}
\caption{Results of the $\zeta$ distribution from predicted posteriors when the MDN-based likelihood is modelled as a single Gaussian distribution. For both the (a) $10\%$ and (b) $0\%$ Dropout MDN-based models, using a single component mixture distribution leads to similar behavior to the $0\%$ BNN prediction from Figure \ref{zeta}.}
\label{hist_gaussian}
\end{figure*}

In addition to the base experiment, we were interested in observing the performance of the MDN-based models when restricted to a single mixture component for the likelihood, essentially modeling it as a simple Gaussian distribution. This allowed us to compare more closely with the single Gaussian posterior predicted using BNNs. We show the results of this experiment in Figure \ref{hist_gaussian}. Interestingly, the $\zeta$ distributions predicted from the MDN-based models show similar poor accuracy as compared to the $0\%$ BNN. This shows that a single Gaussian distribution is not expressive enough to accurately model either the posterior (BNN) or the likelihood (MDN).

In the case of the BNN, we could potentially improve its performance by increasing the number of components. The issue with more components, however, comes from mode collapse, where the network ignores the added number of components in favor of predicting a single component Gaussian distribution. As was done for training mixture density networks, reducing the lensing simulations down to low dimensional statistics to learn a conditional probability distribution helped reduce the risk of mode collapse. A similar framework was used in \cite{2019arXiv190603631M}, where instead of using a CNN to directly output a mixture distribution, the problem is initially separated in two stages: a CNN is first trained to reduce an input image down to low dimensional hypotheses (e.g., means and variances of Gaussian distributions), before training a fully-connected network to predict soft-assignments of the hypothesis used to estimate components of a mixture distribution. The authors noted that separating the inference task into two stages helped reduce the risk of mode collapse for mixture distributions.

\subsection{Computational requirement}

Generating 100 million lensing simulations and obtaining the compressed statistics using our BNN took approximately 300 GPU hours (3 hours across 100 GPUs in parallel). This time is made longer by the fact that we perform 30 forward passes on the same input to perform Monte Carlo Dropout. As such, it is significantly faster when computing compressed statistics from a deterministic neural network, which is what would be preferred for real applications of this work.

Despite the fact that sampling the posterior using an MDN likelihood required running an MCMC on this model, this proved to be extremely fast. Posterior samples for the accuracy tests were obtained by running an MCMC chain for 11000 steps, where the first 10000 steps consisted of a burn-in stage to ensure convergence of the chain. About 200 walkers were used and thinning was done by keeping only 1 sample per 10 steps in order to remove correlations between samples. In total, this gives us 20000 final samples from each of the 100000 posteriors in our test set. The entire sampling process took approximately 20 hours on a single NVIDIA A100 with 40GB of memory. This shows that from a computational cost aspect, the proposed density estimation method is well suited for inferring the posteriors of tens of thousands of upcoming lenses in upcoming surveys.

If even more speed is required, it may be possible to directly model the posterior distribution using MDNs \citep{2019MNRAS.488.4440A} or other types of density estimators \citep[e.g.,][]{2015arXiv150203509G, 2017arXiv170507057P, 2020arXiv201105991J}. This would allow for fast sampling from the posterior distribution, bypassing the need for running an MCMC in the case where the likelihood function is modeled instead. However, this comes at a detriment in flexibility in regards to changing the prior distribution. In this setup, the prior would be implicitly learned by the density estimator modeling the posterior distribution. Therefore, changing the prior during inference would require retraining the density estimator using training data sampled from the new prior distribution (which is also the case for BNNs). This may be suitable for tasks where the prior distribution does not change across different observations. However, for our purposes, given that the prior knowledge of the lensing parameters may be different on a case-by-case basis, training a density estimator to model the posterior for each of the thousands of systems expected to be discovered in upcoming surveys is clearly intractable.

\section{Discussion}
\label{discussion}

The method proposed here breaks the problem of the inference of lensing parameters into two distinct stages: information extraction and posterior modeling. In the first stage, the task of estimating a point estimate directly from high dimensional data, which is a difficult and complex process, is performed efficiently by neural networks, albeit with unknown errors and mistakes. These errors are the result of noise in the data, the degeneracy of parameters, and errors related to the neural networks themselves (e.g., limited expressiveness, poor training, etc.). The second stage uses an independent set of simulations to model the distribution of the true parameters in the simulations and the predictions of the CNN for them. This is a relatively easy density estimation problem in low-dimensional space, which could be done with little error using simple models. It could either be done with traditional density estimation methods \cite[e.g., Gaussian mixture models, ][]{2018MNRAS.477.2874A}, or it could be accelerated by more advanced machine learning methods (e.g., neural networks), as was done in this work. Given that the density modeling captures and quantifies all errors caused by the CNN in the first stage, it minimizes the risk of catastrophic failures; even if the CNN's predictions have significant errors, for example, due to poor training of the network, the posteriors remain accurate, although precision is lost.

This separation of tasks also allows one to independently probe the effect of different factors on the inferred posteriors in a clear manner (e.g., the impact of out-of-distribution training data, the expressiveness of posterior modeling, etc.). For example, to test if the posteriors have more complex structures, one can make the density estimator model more expressive without the need to retrain the point estimate CNN. One could also assess the performance of the model for out-of-distribution data, by changing the simulation set used in training the density model, $\textbf{P}$, to include an additional set of data, $\textbf{R}=\{(\theta_{out}, \hat{\theta}_{out, 1}), (\theta_{out, 1}, \hat{\theta}_{out, 2}), ..., (\theta_{out, M}, \hat{\theta}_{out, M}) \}$, where $\theta_{out}$ represents a model parameter sampled outside of the distribution used for generating the CNN's training data, and $\hat{\theta}_{out}$ is the corresponding CNN prediction.

As noted in \cite{2020arXiv201105991J} and \cite{2022arXiv220300690W}, a simple feed-forward neural density estimator, which produces a parametric distribution (e.g., the mean and the covariance matrix of a multivariate normal distribution) directly from images is also a form of implicit likelihood inference, since the density of the posterior is not explicitly known and is approximated over the training data. This single-stage process, however, has the disadvantage that the complex task of analysing images and extracting relevant information from them and the simpler task of modeling the posterior distributions are combined into a single network. In this form, such a network cannot take into account its own errors for its prediction of the posterior. That is why an approximate Bayesian neural network was used in \cite{2017ApJ...850L...7P} to account for this (epistemic) source of uncertainty. However, the variational approximations make this a less accurate choice and do not provide guarantees in realistic settings.   

A caveat of our analysis here is that the errors of the density estimation models have not been accounted for. In our experiments, however, we found that these errors were quite small. As seen in Figures \ref{cal_default} and \ref{cdf_default}, the posteriors are more accurate than the requirements for most applications. However, if more accuracy is needed, a number of avenues could be explored, including the quantification of density estimation errors. In addition, one could explore the use of more flexible density estimation models such as Masked Autoregressive Flows \citep[MAF,][]{2017arXiv170507057P} for modelling the likelihood of compressed statistics. MAFs are a powerful type of normalizing flow based on a class of density estimation networks known as Masked Autoencoders for Distribution Estimation \citep[MADE][]{2015arXiv150203509G}, which, similar to MDNs, are networks that predict parameters of distributions. The power of MAF lies in its ability to stack multiple layers of MADE networks to allow for highly flexible density models. As such, MAFs could be used to further improve the accuracy of the inferred posterior distributions beyond what is practical with Gaussian mixture models.

As mentioned earlier, another major advantage of the proposed framework is the fact that the priors of the lensing parameters are not burned into the networks and can be imposed on a case-by-case basis at inference time. This could be of significant importance when other probes (e.g., the stellar light of the lens, or the environment of the system) are used to place priors on individual systems. 

This also decouples the prior in the Bayesian analysis framework from the distribution of the training data for the point estimation network. For realistic networks with limited expressivity, the distribution of the parameters in the training set has a major impact on their performance; networks generally perform poorly for data that are encountered less frequently during training. In the proposed method, the distribution of the training data for the CNN can be chosen entirely independently from the Bayesian priors of the parameters. For example, these distributions could be chosen to be flat to allow the CNN to be trained on all regions of the parameter space with equal frequency. 

One drawback of the method proposed here, however, is that modelling the joint posterior of the parameters of interest and of nuisance parameters is necessary in order to marginalize over the nuisance parameters. 
For example, if the source parameters are not modelled during the density estimation stage, the posterior of a particular observed data will be essentially marginalized over the entire prior of the background sources (instead of the posterior) since the MDNs retain no information about the source parameters, resulting in a loss of precision. One possibility for circumventing this issue is to have a separate approximate inference of the nuisance parameters and to draw the simulations for the density estimation stage from these posteriors. This may be possible in a scenario where a separate network is trained to infer a posterior distribution for the nuisance parameters \citep[e.g., a generative model for the background source conditioned on the data, see ][]{Adam:2022b,Adam:2022a}. As mentioned earlier, alternative implicit likelihood inference frameworks that allow implicit marginalization over nuisance parameters (e.g., likelihood ratio methods) are generally only practical in low-dimensional spaces.

This, however, could also be considered a strength of the method, since it allows drawing samples from the full joint posterior of all the parameters (including nuisance parameters) controlling the model to produce simulations that are consistent with the observed data. Methods that implicitly marginalize over nuisance variables (e.g., neural ratio estimators) cannot provide any information about these parameters. Therefore, they do not provide the possibility of verifying their predictions by producing simulations from the inferred posteriors (i.e. models of the data).

In the method proposed here, there is no need for the compressor to take into account its own uncertainty: the second network (MDN) can learn and quantify the compressor network's error.
Despite knowing this, we still used compressed statistics from approximate BNNs as it allows for direct comparison, for the same trained compressor network, between its predicted posteriors and the ones obtained using the MDN model. This comparison was motivated by the fact that BNNs are currently a popular method for predicting uncertainties of neural network predictions. 

\section{Conclusion}
In this work, we proposed a two stage implicit likelihood inference method to obtain the posteriors of macro-lens parameters. In the first stage a compressor network is trained with mean-squared error loss to predict point estimates of the lensing parameters. In the second stage, a density mixture network and a set of simulations are used to model the likelihood of compressed statistics based on repeated simulations of the problem. This is done under a well-defined Bayesian statistical framework that guarantees convergence to the true posterior with reasonably simple requirements and minimizes the risks of catastrophic failures. The model is amortized and can infer the posterior of thousands of systems with limited computational resources. It also allows choosing the prior distribution on a case-by-case basis at inference time.

\begin{acknowledgments}
This research was made possible by a generous donation from Eric and Wendy Schmidt by recommendation of the Schmidt Futures Program. The work was also enabled in part by computational resources provided by Calcul Quebec, Compute Canada and the Digital Research Alliance of Canada. Y.H. and L.P. acknowledge support from the National Sciences and Engineering Council of Canada grant RGPIN-2020-05102, the Fonds de recherche du Québec grant 2022-NC-301305, and the Canada Research Chairs Program. B.D.W. acknowledges support by the ANR BIG4 project, grant ANR-16-CE23-0002 of the French Agence Nationale de la Recherche; and the Labex ILP (reference ANR-10-LABX-63) part of the Idex SUPER, and received financial state aid managed by the Agence Nationale de la Recherche, as part of the programme Investissements d'avenir under the reference ANR-11-IDEX-0004-02. The Flatiron Institute is supported by the Simons Foundation. We thank Adam Coogan and Alexandre Adam for useful discussions in regards to implicit likelihood inference methods.
\end{acknowledgments}
%


\bibliography{bibliography}{}
\bibliographystyle{aasjournal}





\appendix

\section{Bayesian Neural Networks}

\label{bnn}
We train an approximate Bayesian neural network to predict the distribution of the lens and background source parameters from strong lensing images. Approximate BNNs can be used with variational inference to represent the marginalized posterior distribution $p(\theta | x)$ as

\begin{equation}
p(\theta | x) \approx \int p(\theta | x, w) q(w) dw,
\end{equation}

where $q(w)$ is the variational distribution of the network weights. Similar to \cite{2017ApJ...850L...7P}, $q(w)$ is chosen to be a Bernoulli random variable multiplied by the network weights, effectively setting a random number of weights to zero. In practice, this is achieved using Dropout \citep{2015arXiv150602142G} on the output of each network layer except the final layer. The distribution over the target lensing parameters $p(\theta | x, w)$ is chosen to be a multivariate Gaussian distribution, expressed as 

\begin{equation}
    p(\theta | x, w)  = \mathcal{N}\left(\theta; \mu(x, w), \Sigma(x, w)\right),
\end{equation}

where $\mu$ and $\Sigma$ represent the mean and the full covariance matrix. Sampling from the approximate marginalized posterior $p(\theta | x)$ is done by first feeding many times the input $x$ to the BNN in order to get different predictions for $p(\theta | x,w)$, where each prediction is made from a different set of weights $w$ due to random Dropout. Then, data points are drawn from these predicted distributions, forming a set of samples that cover the distribution $p(\theta | x)$.

\subsection{Network Architecture}

The BNN architecture is first composed of a convolutional layer with 32 filters and $7 \times 7$ kernel size, and then four subsequent modules each containing a regular convolution layer with $3 \times 3$ kernels followed by a $1 \times 1$ convolution outputting half the number of filters of the previous layer. This $1 \times 1$ convolution performs pixel-wise connections between the network's feature maps and reduces the complexity of the channel dimension before the next module. The first layer within each module spatially downsamples the feature maps using convolution strides of two. The number of filters used in the first layer of the first module is 64, and is doubled for every subsequent module, reaching a maximum of 512 channel dimensions. Every convolution layer in the network  is followed by a Dropout layer \citep{JMLR:v15:srivastava14a} with equal Dropout rate throughout the network. Besides the final layer, all layers use the Leaky ReLU activation function with a slope of $0.3$ for  negative input values.

Following the convolution modules, the network ends with two fully connected layers; the first having 512 neurons and the second layer outputting the parameters for the distribution $p(\theta | x, w)$. Only the first fully connected layer is followed by a Dropout layer with the same Dropout rate used for the convolutions. While a linear activation function is used for $\mu$ and the off-diagonal elements of the covariance matrix $\Sigma$, the diagonal elements of $\Sigma$ are passed through an exponential activation function to ensure positive values.

\subsection{Training}

Benefiting from the GPU accelerated implementation of our simulation code, new sets of lensing simulations are quickly generated on-the-fly for each forward pass of the BNN during training. The BNN is trained using the Adam optimizer \citep{2014arXiv1412.6980K} with a learning rate of $10^{-4}$ and its weights are optimized for 500000 batches of 32 randomly generated examples each, resulting in an effective training data set of roughly 16 million lensing simulations. As the BNN virtually never encounters the same example twice, the risks of overfitting are minimal. The BNN is trained to minimize the negative log-probability of the predicted distribution as this is equivalent to minimizing the KL divergence between the true and estimated parametric distribution \citep{2019MNRAS.488.4440A}.

After training, the BNN is used to get compressed statistics $\hat{\theta}(x)$ of lensing simulations (by computing the mean of samples from the predicted posteriors), which, along with the true simulation parameters $\theta$, serve as training data for the implicit likelihood inference portion of this work.

\label{additional_results}

\begin{figure}[ht]
  \centering
  \includegraphics[width=1.0\linewidth]{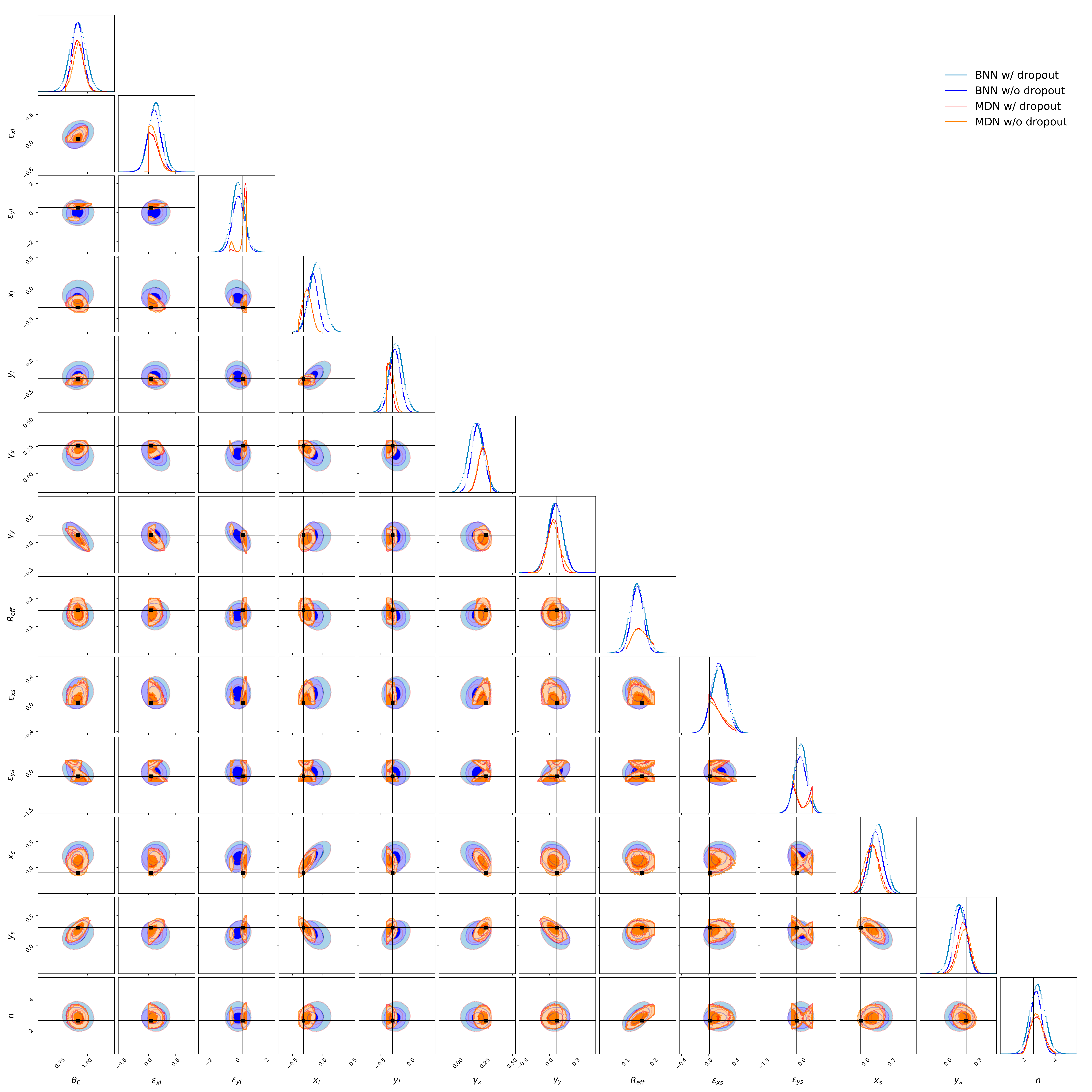}
  \caption{Full posterior of the main lens galaxy and background source parameters from the same observed data used in Figure \ref{degen_post}. A second bimodal distribution is predicted along the $y$ ellipticity of the elliptical background source.}
  \label{degen_post_full}
\end{figure}

\begin{figure}[ht]
  \centering
  \includegraphics[width=0.5\linewidth]{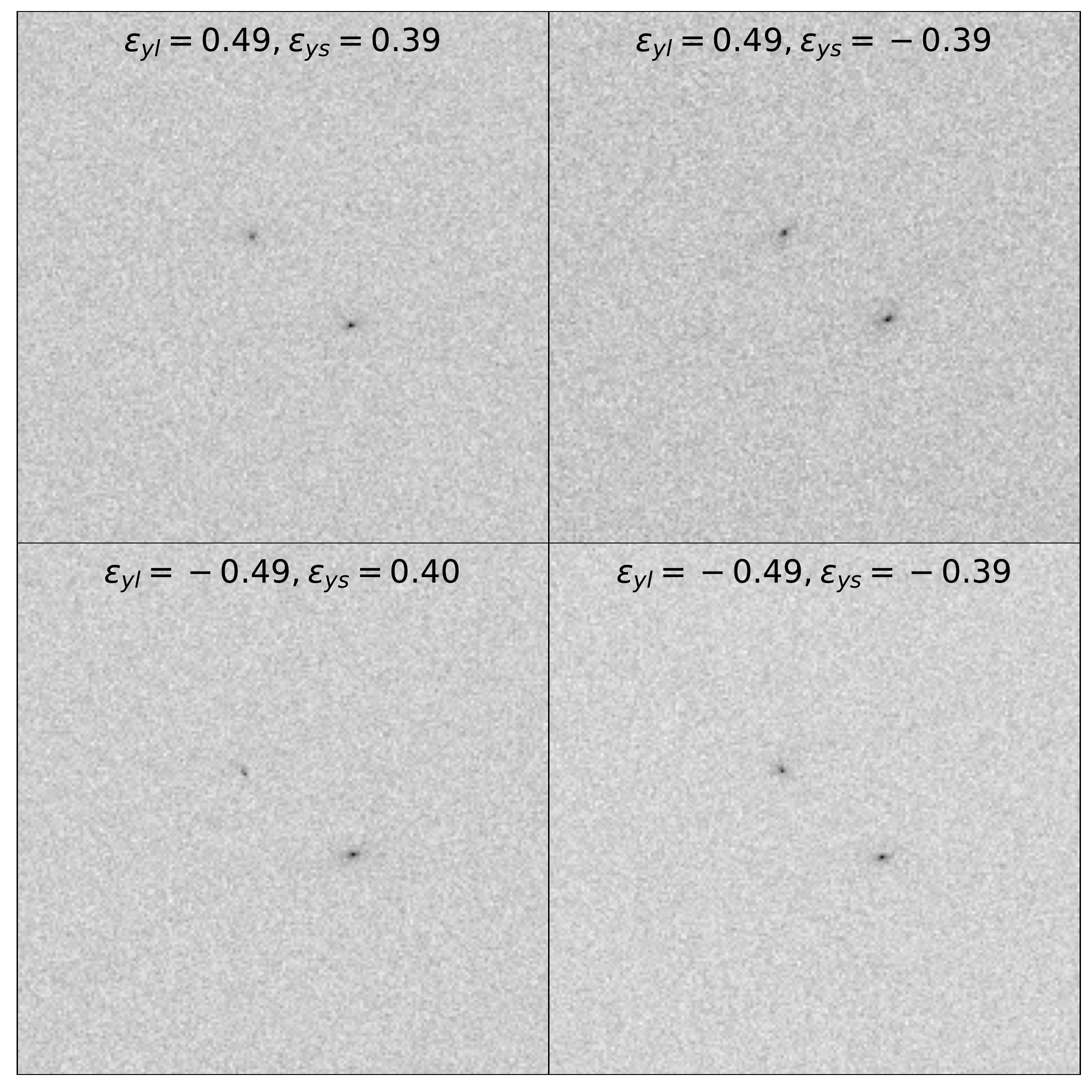}
  \caption{Simulations generated from different pairs of main lens and background source ellipticity with different overall signs from the $10\%$ BNN-based MDN posterior from Figure \ref{degen_post}.}
  \label{mdn0_degens}
\end{figure}

\end{document}